\documentclass[preprint,showpacs,aps,prd,nofootinbib,showkeys,unsortedaddress,raggedbottom]{revtex4-1}
\pdfoutput=1

\usepackage{graphicx}
\usepackage{amsmath}
\usepackage{amssymb}

\usepackage[usenames,dvipsnames]{color}
\definecolor{darkblue}{RGB}{0,0,196}
\definecolor{darkgreen}{RGB}{0,120,0}
\usepackage[colorlinks=true,linkcolor=darkblue,citecolor=darkblue,urlcolor=darkblue]{hyperref}

\def\ba{\begin{eqnarray}}
\def\ea{\end{eqnarray}}

\def\be{\begin{equation}}
\def\ee{\end{equation}}

\newcommand{\checked}[1]{}

\begin{document}

\title{Anisotropic hydrodynamics with a scalar collisional kernel}

\author{Dekrayat Almaalol} 
\author{Michael Strickland} 
\affiliation{Department of Physics, Kent State University, Kent, OH 44242 United States}

\begin{abstract}
Prior studies of non-equilibrium dynamics using anisotropic hydrodynamics have used the relativistic Anderson-Witting scattering kernel or some variant thereof.  In this paper, we make the first study of the impact of using a more realistic scattering kernel.  For this purpose, we consider a conformal system undergoing transversally-homogenous and boost-invariant Bjorken expansion and take the collisional kernel to be given by the leading order $2 \leftrightarrow 2$ scattering kernel in scalar $\lambda \phi^4$.  We consider both classical and quantum statistics in order to assess the impact of Bose enhancement on the dynamics.  We also determine the anisotropic non-equilibrium attractor of a system subject to this collisional kernel.  We find that, when the near-equilibrium relaxation-times in the Anderson-Witting and scalar collisional kernels are matched, the scalar kernel results in a higher degree of momentum-space anisotropy during the system's evolution, given the same initial conditions.  Additionally, we find that taking into account Bose enhancement further increases the dynamically generated momentum-space anisotropy.
\end{abstract}



\maketitle

\section{Introduction}

Understanding the relativistic dynamics of out-of-equilibrium systems is of great importance in both astrophysics and particle physics.  In the context of particle physics, such questions arise, for example, in the study of the high energy-density matter created in ultrarelativistic $AA$, $pA$, and $pp$ collisions \cite{Muller:2012zq,Jacak:2012dx,Schukraft:2017nbn}.  In the astrophysical context, such conditions are created, for example, during the final stages of binary blackhole or neutron star inspiral \cite{2013rehy.book.....R,PhysRevD.71.024035}.  In the study of heavy-ion collisions, one is naturally led to the study of relativistic fluids which are highly momentum-space anisotropic in the local rest frame~\cite{Strickland:2013uga,Jeon:2016uym,Florkowski:2017olj,Alqahtani:2017mhy}.  This momentum-space anisotropy is dynamically generated by the rapid longitudinal expansion of the matter created in high-energy heavy-ion collisions.  Despite these momentum-space anisotropies, it has been found that the evolution of the quark-gluon plasma (QGP) created in heavy-ion collisions is well-described by dissipative hydrodynamics.  This success has been attributed to the existence of an anisotropic non-equilibrium attractor that drives the ``hydrodynamization'' of the system on a sub fm/c timescale in the center of the plasma~\cite{Heller:2015dha,Keegan:2015avk,Romatschke:2017vte,Bemfica:2017wps,Spalinski:2017mel,Strickland:2017kux,Romatschke:2017acs}.   Faced with the existence of an anisotropic dynamical attractor, it is natural to consider fluids that have intrinsic, and potentially large, momentum-space anisotropies.  The framework of anisotropic hydrodynamics (aHydro) was introduced some years ago~\cite{Florkowski:2010cf,Martinez:2010sc} to do just this and has since been extended and applied to QGP phenomenology~\cite{Ryblewski:2010ch,Martinez:2012tu,Ryblewski:2012rr,Bazow:2013ifa,Tinti:2013vba,Nopoush:2014pfa,Tinti:2015xwa,Bazow:2015cha,Strickland:2015utc,Alqahtani:2015qja,Molnar:2016vvu,Molnar:2016gwq,Alqahtani:2016rth,Bluhm:2015raa,Bluhm:2015bzi,Alqahtani:2017jwl,Alqahtani:2017tnq} (for a recent aHydro review see Ref.~\cite{Alqahtani:2017mhy}).

One limitation of all prior aHydro works is their use of the Anderson-Witting collisional kernel~\cite{anderson1974relativistic}, which is otherwise known as the relaxation-time approximation (RTA).  This collisional kernel, while being non-linear due to the Landau matching of the equilibrium and non-equilibrium energy densities, is still conceptually based on a near-equilibrium limit for the collisional kernel.  It is expected that, as the system becomes highly-anisotropic in momentum-space (far from equilibrium), the intrinsic non-linearities in more realistic scattering kernels could become important to the dynamical evolution and the associated non-equilibrium attractor.  In fact, a given collisional kernel can be mapped to an infinite set of transport coefficients in the language of all-order viscous hydrodynamics.  In this paper, we make the first attempt to consider a more realistic scattering kernel in the context of aHydro by considering the leading-order (LO) collisional kernel stemming from \mbox{$2 \leftrightarrow 2$} scattering in massless $\lambda \phi^4$ theory using both classical and quantum (Bose) statistics.  For this conformal theory, it is possible to reduce the necessary ingredients to a finite set of numerically tabulated functions of the momentum-space anisotropy parameter(s) with the scale dependence appearing as an overall multiplicative factor.  In this first work, we consider a transversally homogeneous and boost-invariant system undergoing 0+1d Bjorken expansion and compare to results obtained using RTA.  We demonstrate that the choice of the collisional kernel affects the dynamics quantitatively but not qualitatively.  We further demonstrate that, when the shear relaxation times are matched, the system develops a higher level of momentum-space anisotropy when using the classical scalar kernel than when using RTA.  We also find that incorporating quantum statistics further increases the level of momentum-space anisotropy developed during the evolution.

The structure of the paper is as follows.  In Sec.~\ref{sec:boltz} we introduce the Boltzmann equation and the LO scalar collisional kernel that will be used herein.  In Sec.~\ref{sec:rta} we compute the necessary moments of the collisional kernel in RTA for purposes of comparison with the LO scalar moments.  In Sec.~\ref{sec:matching} we match the LO scalar and RTA moments by requiring that they have the same near-equilibrium relaxation time.  In Sec.~\ref{sec:0+1eqs} we present the general form of the 0+1d aHydro equations of motion that result from taking moments of the Boltzmann equation.  In Sec.~\ref{sec:numericalsolution} we present representative numerical solutions of the aHydro equations of motion, comparing the LO scalar collisional kernel and the RTA collisional kernel.  In Sec.~\ref{sec:attractor} we present the non-equilibrium dynamical attractor emerging from kinetic theory with the LO scalar collisional kernel for both classical and quantum statistics.  In Sec.~\ref{sec:conclusions} we provide our conclusions and an outlook for the future.

\subsection*{Conventions and notation}

Unless otherwise indicated, the Minkowski metric tensor is taken to be ``mostly minus'', i.e. $g^{\mu\nu}={\rm diag}(+,-,-,-)$.  We define the Lorentz-invariant integration measure

\be
\int dP \equiv \int \frac{d^4p}{(2\pi)^4} \, 2\pi \delta(p^\mu p_\mu - m^2) \, 2 \theta(E_p) = \int \frac{d^3{\bf p}}{(2\pi)^3} \frac{1}{E_p}  \, ,
\label{eq:invphase}
\ee
\checked{md}
for a four-vector $p^\mu = (E_p,{\bf p})$.  In what follows, we will work in the massless limit $m \rightarrow 0$ such that $E_p = |{\bf p}|$.

\section{Boltzmann equation with $2 \leftrightarrow 2$ scattering}
\label{sec:boltz}

\begin{figure*}[h!]
\centerline{
\includegraphics[width=.275\linewidth]{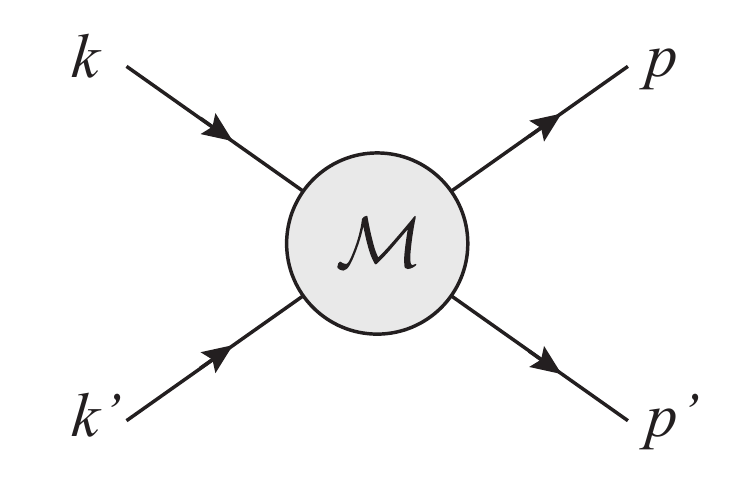}
}
\vspace{-8mm}
\caption{Generic two-to-two scattering diagram.}
\label{fig:2scat}
\end{figure*}

The Boltzmann equation for $2 \leftrightarrow 2$ scattering of identical particles ($k k' \rightarrow p p'$ as depicted in Fig.~\ref{fig:2scat}) is
\be
p^\mu \partial_\mu f_p =  C[f_p] \, ,
\label{eq:boltzmann1}
\ee
\checked{md}
with
\be
C[f_p] = \frac{1}{32} \int dK dK' dP' \, | {\cal M} |^2 \, (2\pi)^4 \delta^{(4)}(k^\alpha + k'^\alpha - p^\alpha - p'^\alpha) \, {\cal F}(k,k',p,p') \, ,
\ee
\checked{md}
where $f_p = f(p)$, etc. and
\be
{\cal F}(k,k',p,p') \equiv f_k f_{k'} (1+a f_p)(1+a f_{p'}) - (1+ a f_k)(1+ a f_{k'}) f_p f_{p'} \, ,
\label{eq:Fdef}
\ee
\checked{md}
with $a=0$ or $1$ for classical or quantum (Bose) statistics, respectively, $k^\alpha$, $k'^\alpha$, $p^\alpha$, and $p'^\alpha$ are understood to be four-vectors of the form $k^\alpha = (E_k,{\bf k})$, etc., and ${\cal M}$ is the transition amplitude.  Although written as four-vectors, all momenta are understood to be on-shell such that, e.g., $E_k = \sqrt{{\bf k}^2+m^2}$.  We will take the massless (conformal) limit, which implies that for on-shell particles $E_k = |{\bf k}| \equiv k$, etc.

We will assume that all distribution functions appearing above are of Romatschke-Strickland (RS) form \cite{Romatschke:2003ms}.  For example,
\be
f_p = f_{\rm eq}\!\left(\frac{1}{\Lambda} \sqrt{p_\perp^2+(1+\xi) ({\bf p}\cdot\hat{\bf n})^2}\right) ,
\label{eq:frs1}
\ee
\checked{md}
with $f_{\rm eq}(x) = 1/[\exp(x)-a]$ with again $a=0$ or 1 for classical and Bose statistics, respectively.
In what follows, we will assume that $\hat{\bf n}$ is a unit-vector along the $z$-direction and hence $p_\perp$ is confined to the $xy$-plane.  Here $-1 < \xi < \infty$ is the anisotropy parameter and $\Lambda$ is the scale parameter.  Both $\xi$ and $\Lambda$ should be understood to be functions of spacetime.  Below we will explicitly consider the case of a transversally homogenous and boost invariant system undergoing Bjorken expansion (0+1d), in which case $\xi$ and $\Lambda$ become functions of only longitudinal proper-time $\tau$.

To proceed, one can take moments of Eq.~(\ref{eq:boltzmann1}) using the integral operator.  
\be
\hat{\cal O}_n \, g = {\cal O}^{\mu_1 \mu_2 \cdots \mu_n}[g]  \equiv \int dP \, p^{\mu_1}p^{\mu_2} \cdots p^{\mu_n} \, g(p) \, .
\ee
\checked{md}
For a general moment of the Boltzmann equation, one obtains
\be
\partial_\mu I^{\mu\nu_1\nu_2\cdots\nu_n} = {\cal C}^{\nu_1\nu_2\cdots\nu_n} \, ,
\label{eq:generaleom}
\ee
\checked{md}
where
\be
I^{\mu\nu_1\nu_2\cdots\nu_n} \equiv \int dP \, p^{\mu} p^{\nu_1} p^{\nu_2} \cdots p^{\nu_n} f \, ,
\ee
\checked{md}
and
\be
{\cal C}^{\nu_1 \nu_2 \cdots \nu_n} \equiv \int dP \, p^{\nu_1}p^{\nu_2} \cdots p^{\nu_n} \, C[f] \, ,
\label{eq:calcdef}
\ee
\checked{md}
are the moments of the distribution function and collisional kernel, respectively.

In number and energy-momentum conserving theories, one finds that the first two moments of the collisional kernel vanish by symmetry, i.e. ${\cal C} = 0$ and ${\cal C}^\mu=0$.  The second moment of the collisional kernel enters into the equation of motion for the third moment of the distribution function
\be
\partial_\lambda I^{\lambda \mu \nu} = {\cal C}^{\mu \nu} \, ,
\ee
with
\be
{\cal C}^{\mu \nu} = \frac{1}{32} \int dK dK' dP dP' \, | {\cal M} |^2 (2\pi)^4 \delta^{4}(k^\alpha + k'^\alpha - p^\alpha - p'^\alpha) {\cal F}(k,k',p,p') p^\mu p^\nu \, .
\label{eq:cmunu1}
\ee
\checked{md}

To apply the four-dimensional delta function, we use Eq.~(\ref{eq:invphase}) to write
\ba
\int dP' (2\pi)^4 \delta^{(4)}(k + k' - p - p') &=& \int d^4p' \, 2\pi \delta(p'^\alpha p'_\alpha) \, 2 \theta(E_{p'}) \delta^{(4)}(k^\alpha + k'^\alpha - p^\alpha - p'^\alpha) \nonumber \\
&=& 4 \pi \delta\Big((k+k'-p)^2 - ({\bf k}+{\bf k}'-{\bf p})^2 \Big) \theta(k+k'-p) \, . \ \ 
\label{eq:simp1}
\ea
\checked{md}

The argument of the delta function has solutions when $k+k'-p = \pm |{\bf k}+{\bf k}'-{\bf p}|$ and the theta function selects the positive solution.  Both solutions obey
\be
(k+k'-p)^2 = ({\bf k}+{\bf k}'-{\bf p})^2 \, .
\ee
\checked{md}
Expanding this, one obtains
\be
k k' - k p - k'p = k k' \cos\theta_{kk'} - k p \cos\theta_{kp} - k' p \cos\theta_{k'p} \, ,
\ee
\checked{md}
where $\theta_{kk'}$ is the relative angle between $k$ and $k'$, etc.  Solving for $p$ gives
\ba
p \rightarrow \tilde{p} &\equiv& \frac{k k' (1 - \cos\theta_{kk'} ) }{ k (1 - \cos\theta_{kp} ) + k' (1 - \cos\theta_{k'p} ) } \, .
\nonumber \\
&=& \frac{k k' - {\bf k}\cdot{\bf k}' }{ k + k' - {\bf k}\cdot\hat{\bf p}  - {\bf k}'\cdot\hat{\bf p} } \, ,
\label{eq:psol1}
\ea
\checked{md}
where $\hat{\bf p} = {\bf p}/p$.  Note that from above one finds $\tilde{p} \geq 0$.

Therefore, using the general rule for a delta function of a function in Eq.~(\ref{eq:simp1}) gives
\be
\int dP' \, (2\pi)^4 \delta^{(4)}(k^\alpha + k'^\alpha- p^\alpha - p'^\alpha) = \frac{2\pi}{E_{p'}} \delta(p-\tilde{p}) \, .
\ee
\checked{md}
where $E_{p'} = p' = k + k'  - p$ and $\tilde{p}$ is defined in Eq.~(\ref{eq:psol1}).  Inserting this relation into Eq.~(\ref{eq:cmunu1}), one obtains
\be
{\cal C}^{\mu \nu} = \frac{1}{128\pi^2} \int dK dK' d\Omega_p \frac{ p | {\cal M} |^2}{ E_{p'}} {\cal F}(k,k',p,p') p^\mu p^\nu \Bigg|_{p \rightarrow \tilde{p}} \, .
\label{eq:cmunu2}
\ee
\checked{md}
This equation holds for any energy-momentum conserving $2 \leftrightarrow 2$ scattering.  In what follows, we will specialize to the case of LO $2 \leftrightarrow 2$ scattering in $\lambda \phi^4$ theory, in which case $|{\cal M}|^2 = \lambda^2$.  In general, Eq.~(\ref{eq:cmunu2}) is a function of $\xi$ and $\Lambda$, however, in conformal (massless) theories the scale $\Lambda$ can be pulled out by rescaling the momenta, resulting in an overall factor of $\Lambda^6$.  The remaining eight-dimensional integral is then a function only of $\xi$ and can be evaluated using Monte-Carlo integration.

\section{Moments of the RTA collisional kernel}
\label{sec:rta}

All previous results in the context of anisotropic hydrodynamics have assumed that the collisional kernel is given by the relativistic Anderson-Wittig \cite{Anderson:1974} model, which is otherwise known as the ``relaxation-time approximation'' (RTA).  Since we will compare to results obtained using the RTA, it is necessary to relate the scalar coupling constant $\lambda$ and the relaxation time $\tau_{\rm eq}$ appearing in RTA in order to make an apples-to-apples comparison.  In this section, we provide the RTA results.  In the next section, we use these results to match the collisional kernels by requiring that the relaxation time is the same in each theory in the near-equilibrium limit.

The RTA collisional kernel is
\be
C_{\rm RTA}[f_p] = \frac{E_p}{\tau_{\rm eq}} \left[ f_{\rm eq}(p/T) - f_p \right] ,
\ee
\checked{md}
where the four-momentum is specified in the fluid local rest frame, $\tau_{\rm eq} = 5 \bar\eta/T$ with $\bar\eta \equiv \eta/s$ \cite{Denicol:2010xn,Denicol:2011fa}, and \mbox{$T = R^{1/4}(\xi) \Lambda$} \cite{Martinez:2009mf,Martinez:2009ry,Martinez:2010sc} with
\be
{\cal R}(\xi) = \frac{1}{2}\left[\frac{1}{1+\xi}
+\frac{\arctan\sqrt{\xi}}{\sqrt{\xi}} \right] ,
\label{eq:rfunc}
\ee
\checked{md}
\cite{Rebhan:2008uj,Martinez:2010sc}.

The resulting second-moment of the collisional kernel is
\be
{\cal C}^{\mu\nu}_{\rm RTA} = \int dP \, C_{\rm RTA}[f] \, p^\mu p^\nu = \frac{1}{\tau_{\rm eq}} \int \frac{d^3p}{(2\pi)^3} \left[ f_{\rm eq}(p/T) - f_p \right] p^\mu p^\nu \, .
\ee
\checked{md}
In 0+1d case one has ${\cal C}^{xx} = {\cal C}^{yy}$ and all off-diagonal components vanish.  Additionally, $C^{00} = \sum_{i\in\{x,y,z\}} C^{ii}$ since $m=0$.  As a result, there are only two independent components ${\cal C}^{xx}$ and ${\cal C}^{zz}$.  Focusing first on ${\cal C}^{zz}$ since the same method can be used to obtain ${\cal C}^{xx}$, using Eq.~(\ref{eq:frs1}) one obtains
\ba
{\cal C}^{zz}_{\rm RTA} &=&  \frac{1}{\tau_{\rm eq}}  \int \frac{d^3p}{(2\pi)^3} \left[ f_{\rm eq}(p/T) - (1+\xi)^{-3/2} f_{\rm eq}(p/\Lambda) \right] p_z^2 \nonumber\\
&=&  \frac{\Lambda^6}{5 \bar\eta} \left[ {\cal R}^{3/2}(\xi) -  \frac{{\cal R}^{1/4}(\xi)}{(1+\xi)^{3/2}} \right] \int \frac{d^3\bar{p}}{(2\pi)^3}  \bar{p}_z^2  f^{\rm eq}(\bar{p})  \, .
\ea
\checked{m}

For a Boltzmann distribution, one obtains
\be
\kappa_0 \equiv \int \frac{d^3\bar{p}}{(2\pi)^3}  \exp(-\bar{p}) \bar{p}_z^2 = \frac{1}{3} \int \frac{d^3\bar{p}}{(2\pi)^3}  \exp(-\bar{p}) \bar{p}^2 = \frac{4}{\pi^2} \, .
\ee
\checked{md}
and, for a Bose distribution, one obtains
\be
\kappa_1 = \frac{1}{3} \int \frac{d^3\bar{p}}{(2\pi)^3}  \frac{\bar{p}^2}{\exp(\bar{p})-1} = \frac{4\zeta(5)}{\pi^2} \, ,
\ee
\checked{m}
where $\zeta(x)$ is the Riemann zeta function.

\subsection*{Final result}

The final RTA result for ${\cal C}^{zz}$ is
\be
{\cal C}^{zz}_{\rm RTA} = \frac{\kappa_a\Lambda^6}{5\bar\eta} \left[ {\cal R}^{3/2}(\xi) -  \frac{{\cal R}^{1/4}(\xi)}{(1+\xi)^{3/2}} \right]  .
\ee
\checked{md}
The $xx$-projection can be obtained similarly
\be
{\cal C}^{xx}_{\rm RTA} = \frac{\kappa_a\Lambda^6}{5\bar\eta} \left[  {\cal R}^{3/2}(\xi) -  \frac{ {\cal R}^{1/4}(\xi)}{(1+\xi)^{1/2}} \right]  .
\ee
\checked{m}
In both expressions above
\be
\kappa_a = \begin{cases}
    \; \frac{4}{\pi^2} & \quad \text{if } a = 0 \;\; ({\rm classical}) \, ,  \\
    \; \frac{4\zeta(5)}{\pi^2} & \quad \text{if } a = 1 \;\; ({\rm quantum}) \, .
\end{cases}
\label{eq:kapparel}
\ee
\checked{m}

\section{Matching between the scalar and RTA kernels}
\label{sec:matching}

Next, we perform a small anisotropy expansion of the scalar $2 \leftrightarrow 2$ collisional kernel for both classical and quantum statistics and match to RTA by requiring that the relaxation time in each theory is the same in this limit.  To begin we require the small-$\xi$ expansions of the scalar collisional kernel.

\subsection{Classical statistics -- $a=0$}

In the limit $\xi \rightarrow 0$, using Eq.~(\ref{eq:Fdef}) with $a=0$, one has
\be
{\cal F}(k,k',p,p') =  \frac{e^{-\frac{k+{k'}}{\Lambda }}}{2 \Lambda p'} \, {\cal G}({\bf k},{\bf k}',{\bf p},{\bf p}')  \, \xi + {\cal O}(\xi^2) \, ,
\label{eq:smallxi1}
\ee
\checked{md}
with
\ba
{\cal G}({\bf k},{\bf k}',{\bf p},{\bf p}') &=& 
2 k \cos{\theta_{k}} ({k'} \cos {\theta_{k'}}- p \cos {\theta_{p}})
\nonumber \\ && \hspace{1cm}
+\ k (p-{k'}) \cos^2\!\theta_{k}+{k'} (p-k) \cos ^2\!\theta_{k'} \nonumber 
\\ && \hspace{2cm}
+\ p (k+{k'}) \cos ^2\!\theta_{p}  -2 {k'} p \cos{\theta_{k'}} \cos{\theta_{p}}  \, .
\label{eq:calg}
\ea
\checked{md}

\subsection{Quantum statistics -- $a=1$}

In the limit $\xi \rightarrow 0$, using Eq.~(\ref{eq:Fdef}) with $a=1$, one has
\be
{\cal F}(k,k',p,p') =  \frac{e^{\frac{k+{k'}}{\Lambda }} f_{\rm eq}(k/\Lambda) f_{\rm eq}(k'/\Lambda)f_{\rm eq}(p/\Lambda) f_{\rm eq}(p'/\Lambda) }{2 \Lambda p'}  \, {\cal G}({\bf k},{\bf k}',{\bf p},{\bf p}')  \, \xi + {\cal O}(\xi^2) \, ,
\label{eq:smallxi2}
\ee
\checked{md}
with ${\cal G}$ once again given by Eq.~(\ref{eq:calg}).

\subsection{Matching to RTA}

Plugging the leading-order terms listed in Eqs.~(\ref{eq:smallxi1}) or (\ref{eq:smallxi2}) into Eq.~(\ref{eq:cmunu2}) gives an eight-dimensional integral for the small $\xi$ limit for the case of classical and quantum statistics, respectively.  The resulting integrals can be performed numerically using Monte Carlo integration.  For this purpose, we used the GNU Scientific Library (GSL) VEGAS algorithm \cite{gsl}  with $10^7$ evaluations per iteration.  We terminated the iterations when the $\chi^2$ value of the last iteration fell in the range $0.5 < \chi^2 < 1.5$. The results obtained were
\be
\lim_{\xi \rightarrow 0} \frac{{\cal C}^{zz}}{\Lambda^6} = \alpha_a \lambda^2 \xi + {\cal O}(\xi^2) \, ,
\label{eq:smallxi3}
\ee
\checked{md}
with $\alpha_0 \simeq 0.4394 \pm 0.0002$ for classical statistics ($a=0$) and $\alpha_1 \simeq 0.7773 \pm 0.0008$ for quantum statistics ($a=1$).  Note also that one can show that $\lim_{\xi \rightarrow 0} {\cal C}^{xx} = -\frac{1}{2} \lim_{\xi \rightarrow 0} {\cal C}^{zz}$.

Using the results presented in the previous section for ${\cal C}^{zz}_{\rm, RTA}$, one finds
\be
\lim_{\xi \rightarrow 0} \frac{{\cal C}^{zz}_{\rm RTA}}{\Lambda^6} = \frac{2\kappa_a}{15\bar\eta} \xi + {\cal O}(\xi^2) \, ,
\ee
\checked{m}
where $\bar\eta = \eta/s$.  Equating the scalar collisional kernel result, one obtains
\be
\bar\eta = \frac{2 \kappa_a}{15 \alpha_a \lambda^2} \, .
\label{eq:match}
\ee 
\checked{m}
We note, in closing, that the above relation can be used to determine the value of the coupling constant $\lambda$ necessary to achieve a given value of $\bar\eta$ in each case.  When generating our numerical comparisons, we will use Eq.~(\ref{eq:match}) to fix $\lambda$ in order make the value of $\bar\eta$ the same in all cases considered.

\begin{figure*}[t!]
\centerline{
\includegraphics[width=.475\linewidth]{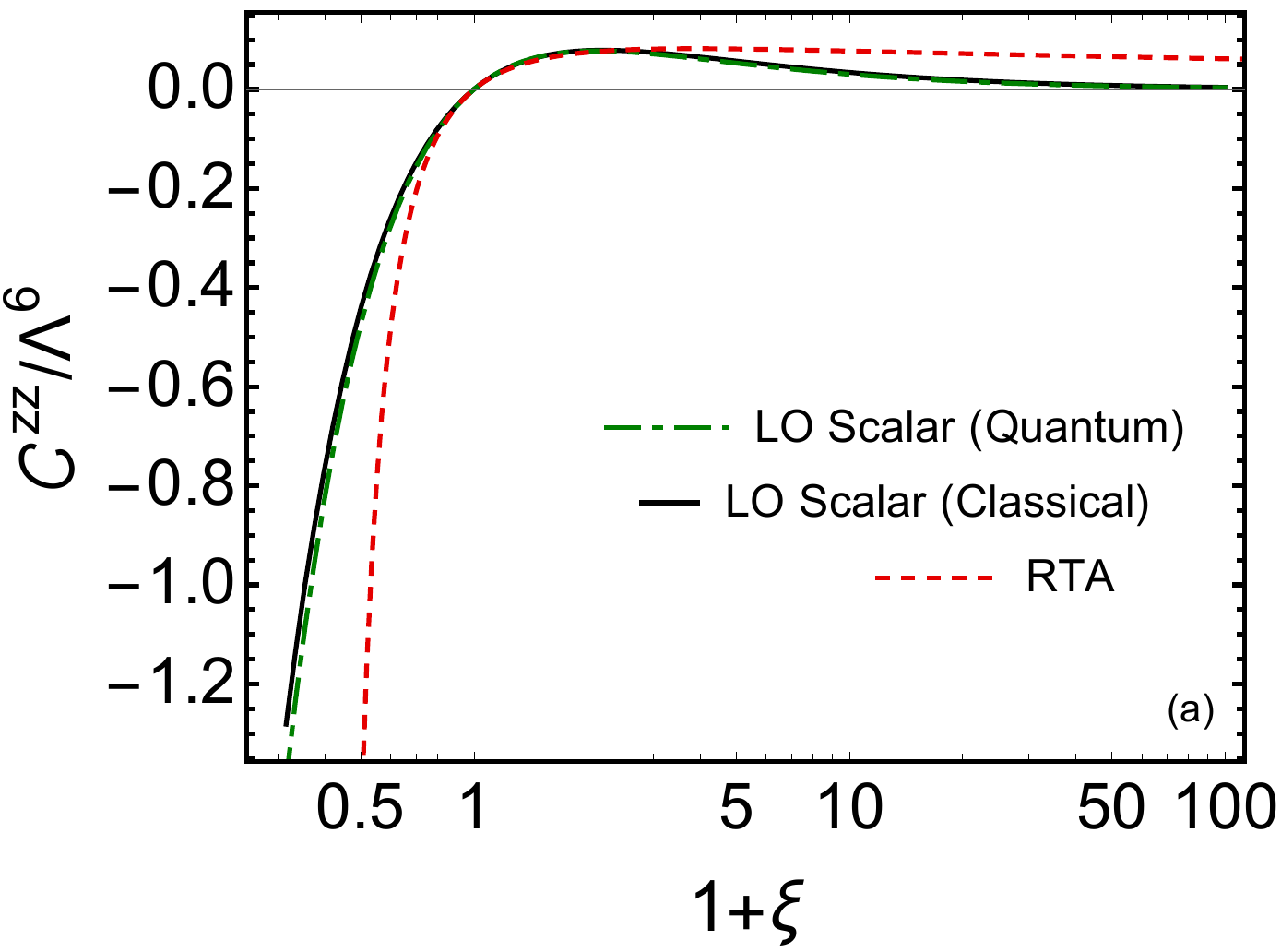}
\hspace{3mm}
\includegraphics[width=.475\linewidth]{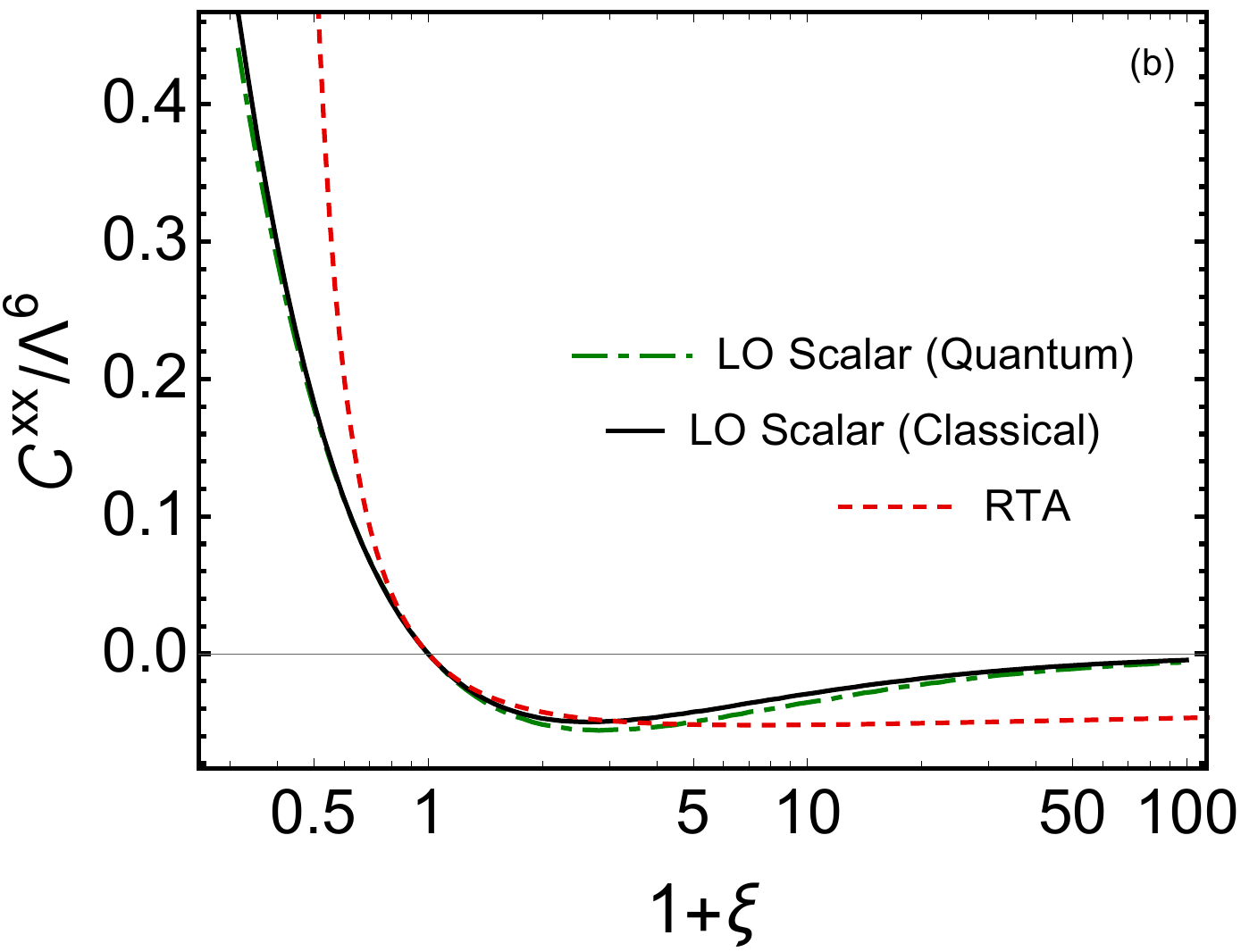}
}
\caption{(Color online) Comparison of the LO scalar scattering kernel moments for both the classical and quantum cases with those obtained in RTA as a function of $\xi$.  Panel (a) shows ${\cal C}^{zz}/\Lambda^6$ and panel (b) shows ${\cal C}^{xx}/\Lambda^6$.  For the purposes of this figure, we took $\bar\eta = 0.2$.}
\label{fig:ccomp1}
\end{figure*}

\begin{figure*}[t!]
\centerline{
\includegraphics[width=.475\linewidth]{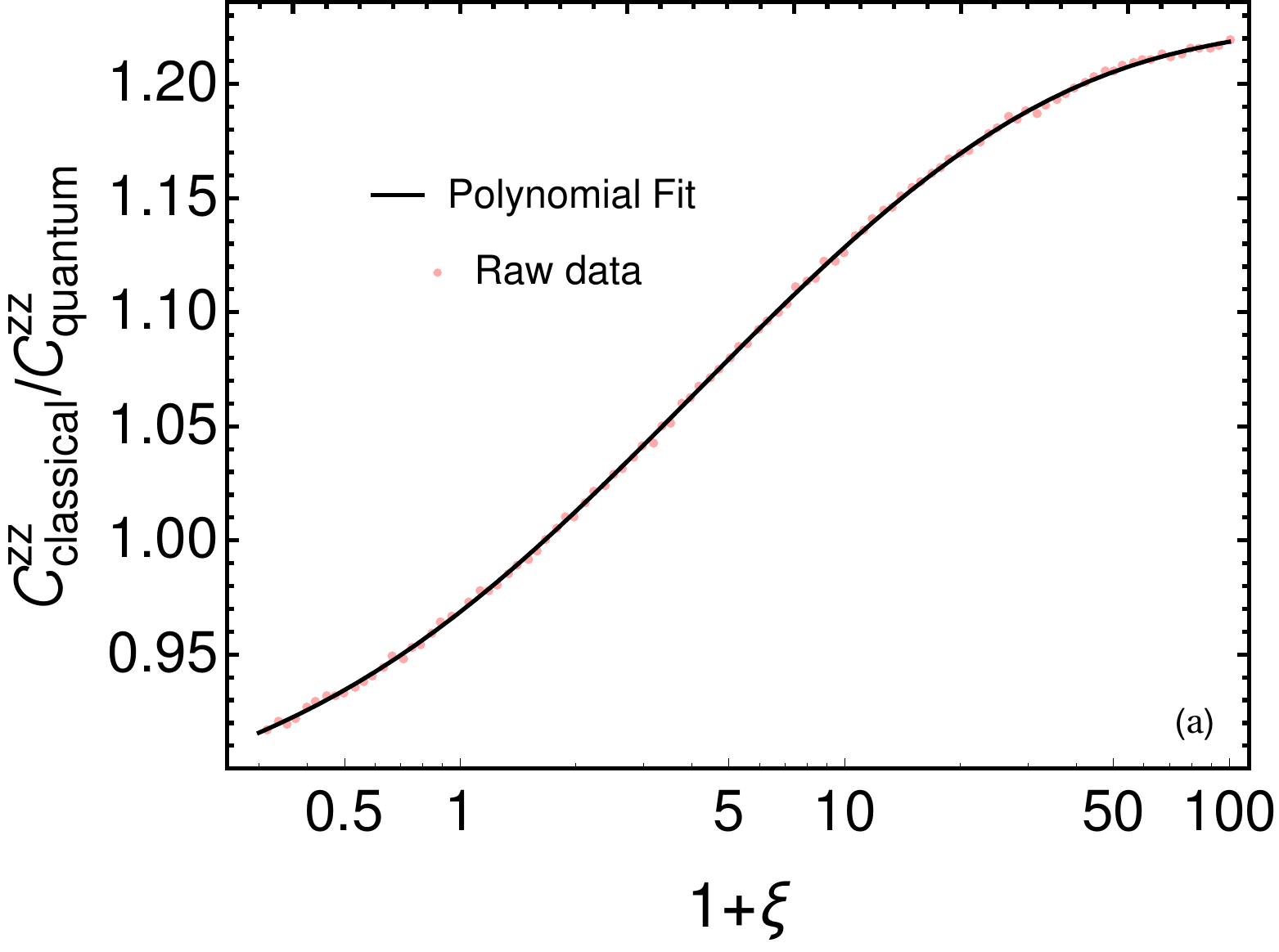}
\hspace{3mm}
\includegraphics[width=.475\linewidth]{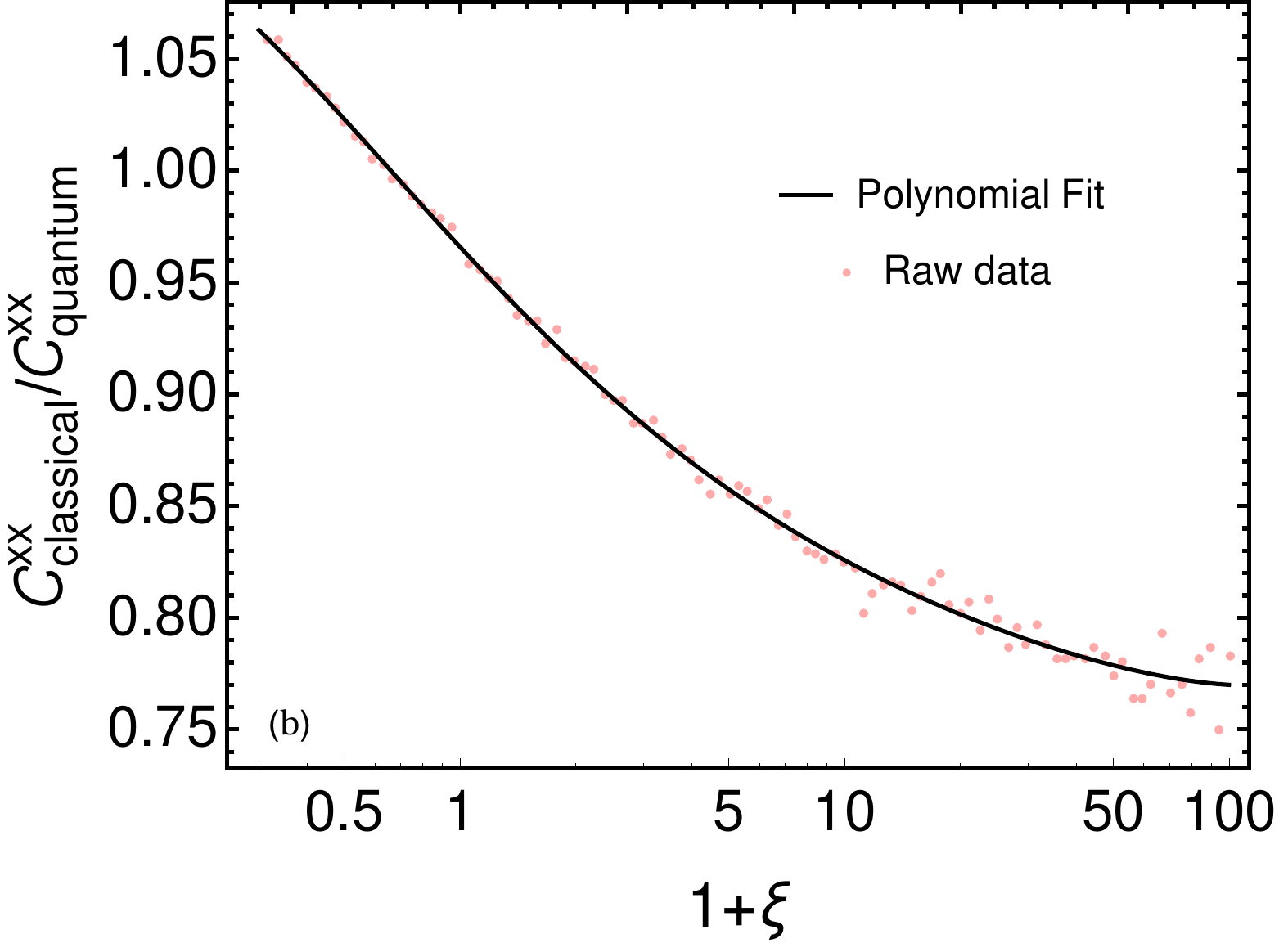}
}
\caption{(Color online) Ratio of the moments of the LO scalar collisional kernel in the classical and quantum and quantum cases as a function of $\xi$. Panel (a) shows ${\cal C}^{zz}_{\rm classical}/{\cal C}^{zz}_{\rm quantum}$ and panel (b) shows ${\cal C}^{xx}_{\rm classical}/{\cal C}^{xx}_{\rm quantum}$.  The red points show the ratio of the two Monte-Carlo results and the solid black line is a fifth-order polynomial fit to the numerical data.}
\label{fig:ccomp2}
\end{figure*}

\subsection{Comparison between the matched scalar and RTA kernel moments}

In Fig.~\ref{fig:ccomp1} we compare ${\cal C}^{zz}$ and ${\cal C}^{xx}$ obtained using the LO scalar collisional kernel and RTA.  For the numerical evaluation of the integrals necessary we again used GSL Monte-Carlo VEGAS with the same number of points per iteration and convergence criteria as in the small-$\xi$ case.  Panel (a) shows ${\cal C}^{zz}/\Lambda^6$ and panel (b) shows ${\cal C}^{xx}/\Lambda^6$.  For the purposes of this figure, we took $\bar\eta = 0.2$ which only affects the results for ${\cal C}^{zz}$ and ${\cal C}^{xx}$ as an overall multiplicative factor which is the same for all kernels.  As can be seen from this figure, due to the matching, all curves coincide in the limit of small anisotropy parameter; however, we observe that both the classical and quantum LO scalar kernel moments are suppressed relative to the RTA result at large values of $\xi$.  One can expect, based on this, that if the system develops an oblate ($\xi>0$) momentum anisotropy, the LO scalar kernel will be less efficient at restoring isotropy and hence a higher degree of oblate momentum-space anisotropy will develop.  A similar conclusion can be drawn from the prolate region ($-1<\xi<0$) where, once again, we see that the magnitude of the RTA moment always exceeds that of the corresponding LO scalar kernel moment.

From Fig.~\ref{fig:ccomp1} we also see that the classical ($a=0$) and quantum ($a=1$) versions of the LO scalar collisional kernel give results which are very close.  To further quantify the difference between these two cases, in Fig.~\ref{fig:ccomp2} we present the ratio of the classical to quantum results for ${\cal C}^{zz}$ and ${\cal C}^{xx}$ in the panels (a) and (b), respectively.  As we can see from this figure, in the range of $\xi$ shown, the difference between the classical and quantum kernel moments is at most approximately 25\%.

\section{0+1d equations of motion}
\label{sec:0+1eqs}

In this section, we derive the conformal 0+1d equations of motion using both the LO scalar and RTA collisional kernels.  In all cases shown, we will use the $uu$ projection of the first moment and the $zz - \frac{1}{3}(xx+yy+zz)$ projection of the second moment to obtain the necessary 0+1d equations of motion.  The first moment equation is independent of the collisional kernel and can be expressed compactly as
\be
\partial_\tau \varepsilon = -\frac{\varepsilon + P_L}{\tau} \, ,
\label{eq:1stmoment}
\ee
\checked{md}
where $\varepsilon = {\cal R}(\xi) \varepsilon_{\rm eq}(\Lambda)$ is the energy density and $P_L = {\cal R}_L(\xi) P_{\rm eq}(\Lambda)$ is the longitudinal pressure with ${\cal R}(\xi)$ defined in Eq.~(\ref{eq:rfunc}) and
\be
{\cal R}_{L}(\xi) = \frac{3}{\xi} \left[ \frac{(\xi+1){\cal R}(\xi)-1}{\xi+1}\right] .
\ee
\checked{md}
We note that, since we consider a conformal system, one has $\varepsilon_{\rm eq}(\Lambda) = 3 P_{\rm eq}(\Lambda)$.

The $xx$, $yy$, and $zz$ projections of the second moment of the Boltzmann equation give \cite{Tinti:2013vba,Nopoush:2014pfa}
\be
\partial_\tau I_i + (\theta - 2 \theta_i) = {\cal C}^{ii} \, ,
\ee
\checked{md}
where $i \in \{x,y,z\}$, $I_i  \equiv u^\mu X_i^\nu X_i^\lambda I_{\mu\nu\lambda}$, and ${\cal C}^{ii} \equiv X_i^\mu X_i^\nu {\cal C}_{\mu\nu}$.  For 0+1d Bjorken expansion one has $\theta=1/\tau$, $\theta_x = \theta_y=0$, and $\theta_z=-1/\tau$.

In an isotropic system, one finds $I_x = I _y = I_z = I_0$ where
\be
I_0(\Lambda) = \kappa_a \Lambda^5 \, ,
\ee
\checked{md}
with $\kappa_a$ given by Eq.~(\ref{eq:kapparel}).  Using the spheroidal aHydro distribution function (\ref{eq:frs1}) one finds
\ba
I_x &=& I_y = {\cal S}_T(\xi) I_0(\Lambda) \, , \nonumber \\
I_z &=& {\cal S}_L(\xi) I_0(\Lambda) \, ,
\ea
\checked{md}
with
\ba
{\cal S}_T(\xi) &=& \frac{1}{\sqrt{1+\xi}} \, , \nonumber \\
{\cal S}_L(\xi) &=& \frac{1}{(1+\xi)^{3/2}} \, .
\ea
\checked{md}

From the $zz$ projection one obtains
\be
(\ln {\cal S}_L)' \partial_\tau \xi + 5 \partial_\tau \ln\Lambda + \frac{3}{\tau} = \frac{{\cal C}^{zz}}{I_{z}}  \, ,
\ee
\checked{md}
and from the $xx$ and $yy$ projections one obtains
\be
(\ln {\cal S}_T)' \partial_\tau \xi + 5 \partial_\tau \ln\Lambda + \frac{1}{\tau} = \frac{{\cal C}^{xx}}{I_{x}}  \, ,
\ee
\checked{md}

Using the fact that $I_z - (I_x+I_y+I_z)/3 = 2(I_z-I_x)/3$ for a 0+1d system, after simplification, the equation of motion necessary becomes
\be
\frac{1}{1+\xi} \partial_\tau \xi  - \frac{2}{\tau} = {\cal C} \, ,
\ee
\checked{md}
where
\be
{\cal C} \equiv \frac{{\cal C}^{xx}}{I_{x}} -  \frac{{\cal C}^{zz}}{I_{z}}  = \frac{\Lambda}{\kappa_a} \left[ (1+\xi)^{1/2} \bar{\cal C}^{xx}(\xi)  - (1+\xi)^{3/2} \bar{\cal C}^{zz}(\xi) \right]  ,
\ee
\checked{md}
and $\bar{\cal C}^{xx}(\xi) \equiv {\cal C}^{xx}/\Lambda^6$ and $\bar{\cal C}^{zz}(\xi) \equiv {\cal C}^{zz}/\Lambda^6$ are dimensionless functions of $\xi$.  This gives our final second moment equation for a general collisional kernel
\be
\frac{1}{1+\xi} \partial_\tau \xi  - \frac{2}{\tau} = \frac{\Lambda}{\kappa_a} \left[ (1+\xi)^{1/2} \bar{\cal C}^{xx}(\xi)  - (1+\xi)^{3/2} \bar{\cal C}^{zz}(\xi) \right] .
\ee
\checked{md}

\subsection{Equations of motion in RTA}

In RTA, one has
\be
\bar{\cal C}^{zz}_{\rm RTA} =  \frac{\kappa_a}{5 \bar\eta} \left[ {\cal R}^{3/2}(\xi) -  \frac{{\cal R}^{1/4}(\xi)}{(1+\xi)^{3/2}} \right]  .
\ee
\checked{md}
Following a similar procedure, ${\cal C}^{xx}_{\rm RTA}$ is found to be
\be
\bar{\cal C}^{xx}_{\rm RTA} =  \frac{\kappa_a}{5 \bar\eta} \left[  {\cal R}^{3/2}(\xi) -  \frac{ {\cal R}^{1/4}(\xi)}{(1+\xi)^{1/2}} \right]  .
\ee
\checked{md}
This gives
\be
{\cal C} =  -\frac{\Lambda}{5\bar\eta} \xi \sqrt{1+\xi} {\cal R}^{3/2}(\xi) \, ,
\ee
\checked{m}
and the resulting dynamical equation is 
\be
\frac{1}{1+\xi} \partial_\tau \xi  - \frac{2}{\tau} + \frac{\Lambda}{5\bar\eta} \xi \sqrt{1+\xi} {\cal R}^{3/2}(\xi) = 0  \, .
\ee
\checked{m}
This agrees with Eq.~(15) of Ref.~\cite{Strickland:2017kux}.

\subsection{Further simplification for scalar $2 \leftrightarrow 2$ scattering}

Introducing $\tilde{C}^{ii} = \bar{\cal C}^{ii}/\lambda^2$ we have
\be
\frac{1}{1+\xi} \partial_\tau \xi  - \frac{2}{\tau} = \frac{\Lambda\lambda^2}{\kappa_a} \left[ (1+\xi)^{1/2} \tilde{\cal C}^{xx}(\xi)  - (1+\xi)^{3/2} \tilde{\cal C}^{zz}(\xi) \right] ,
\ee
\checked{m}
and, using the matching condition (\ref{eq:match}), one has
\be
\lambda^2 = \frac{2 \kappa_a}{15 \alpha_a \bar\eta} \, .
\ee
\checked{m}
Using this, one obtains
\be
\frac{1}{1+\xi} \partial_\tau \xi  - \frac{2}{\tau} = \frac{2\Lambda}{15 \alpha_a \bar\eta} \left[ (1+\xi)^{1/2} \tilde{\cal C}^{xx}(\xi)  - (1+\xi)^{3/2} \tilde{\cal C}^{zz}(\xi) \right] .
\label{eq:2mom1}
\ee
\checked{m}

\begin{figure*}[t!]
\centerline{
\includegraphics[width=.475\linewidth]{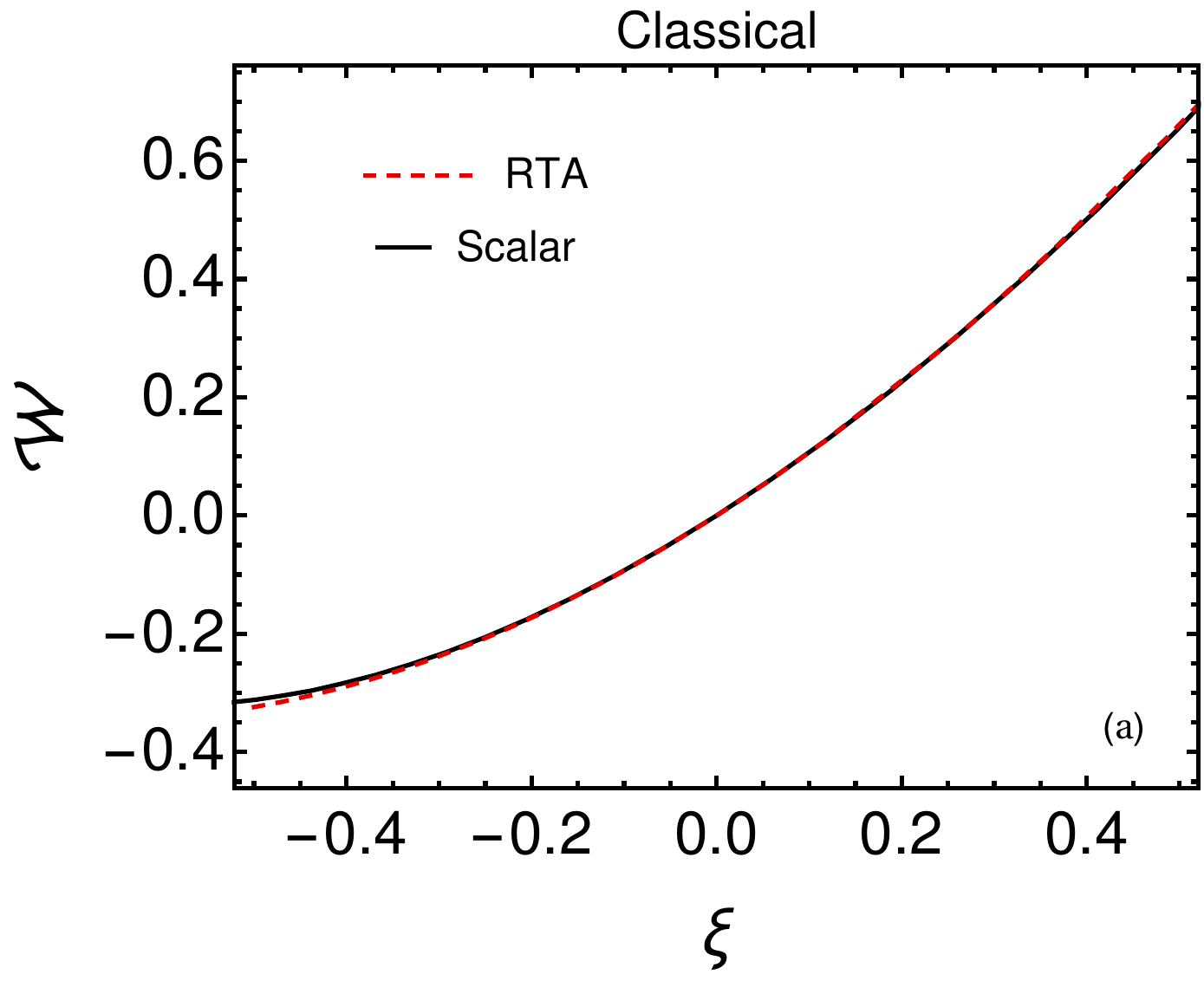}
\hspace{3mm}
\includegraphics[width=.475\linewidth]{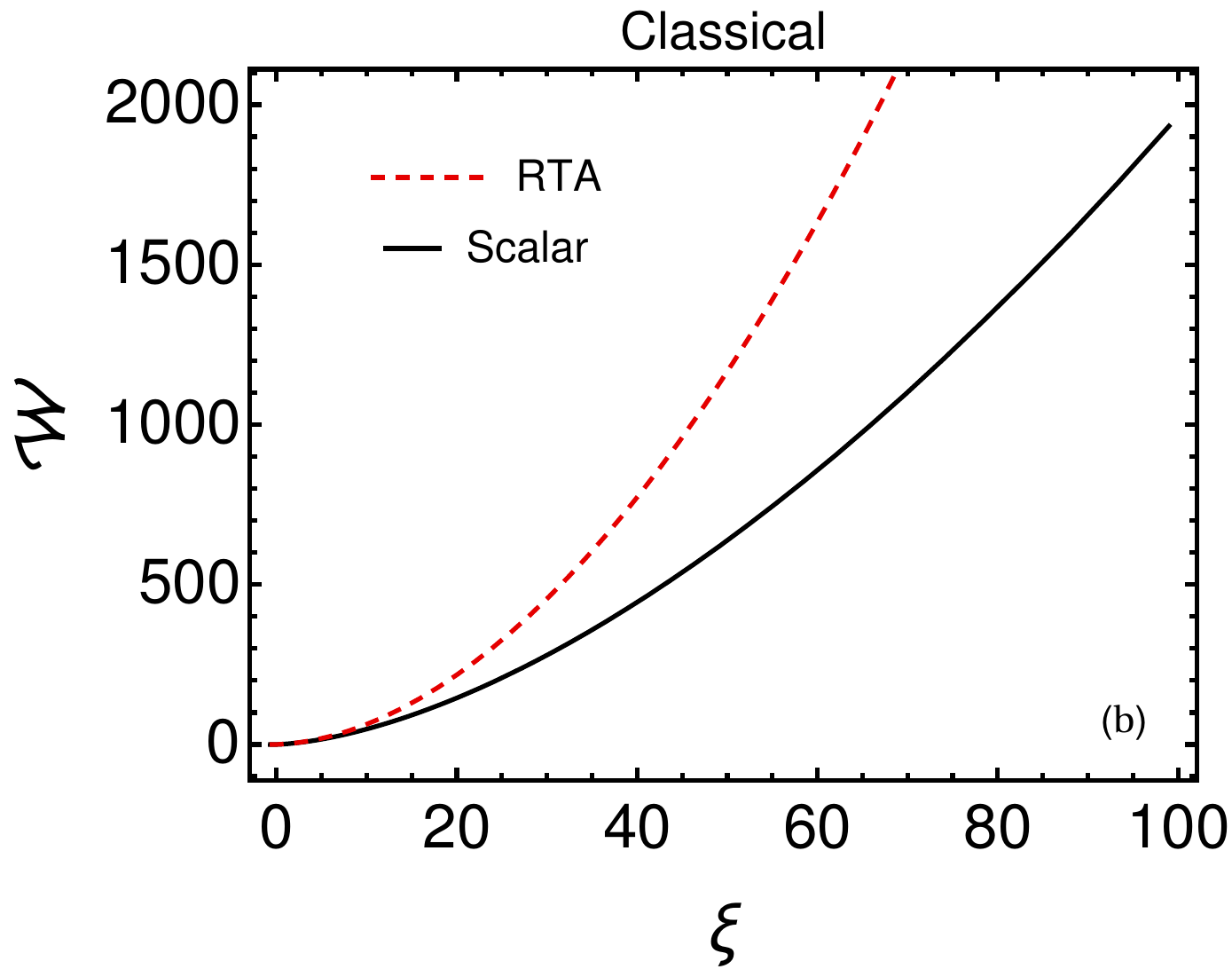}
}
\vspace{4mm}
\centerline{
\includegraphics[width=.475\linewidth]{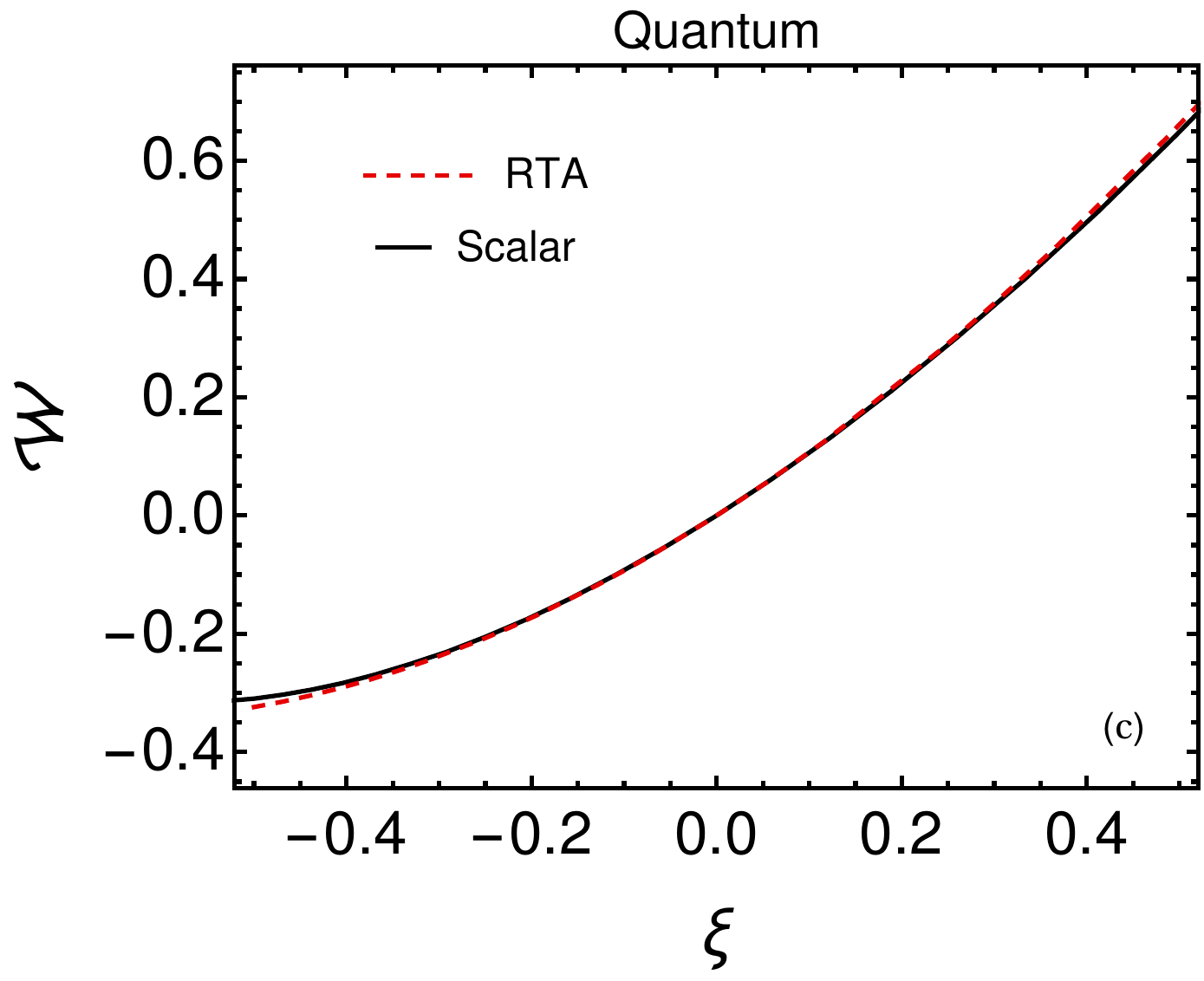}
\hspace{3mm}
\includegraphics[width=.475\linewidth]{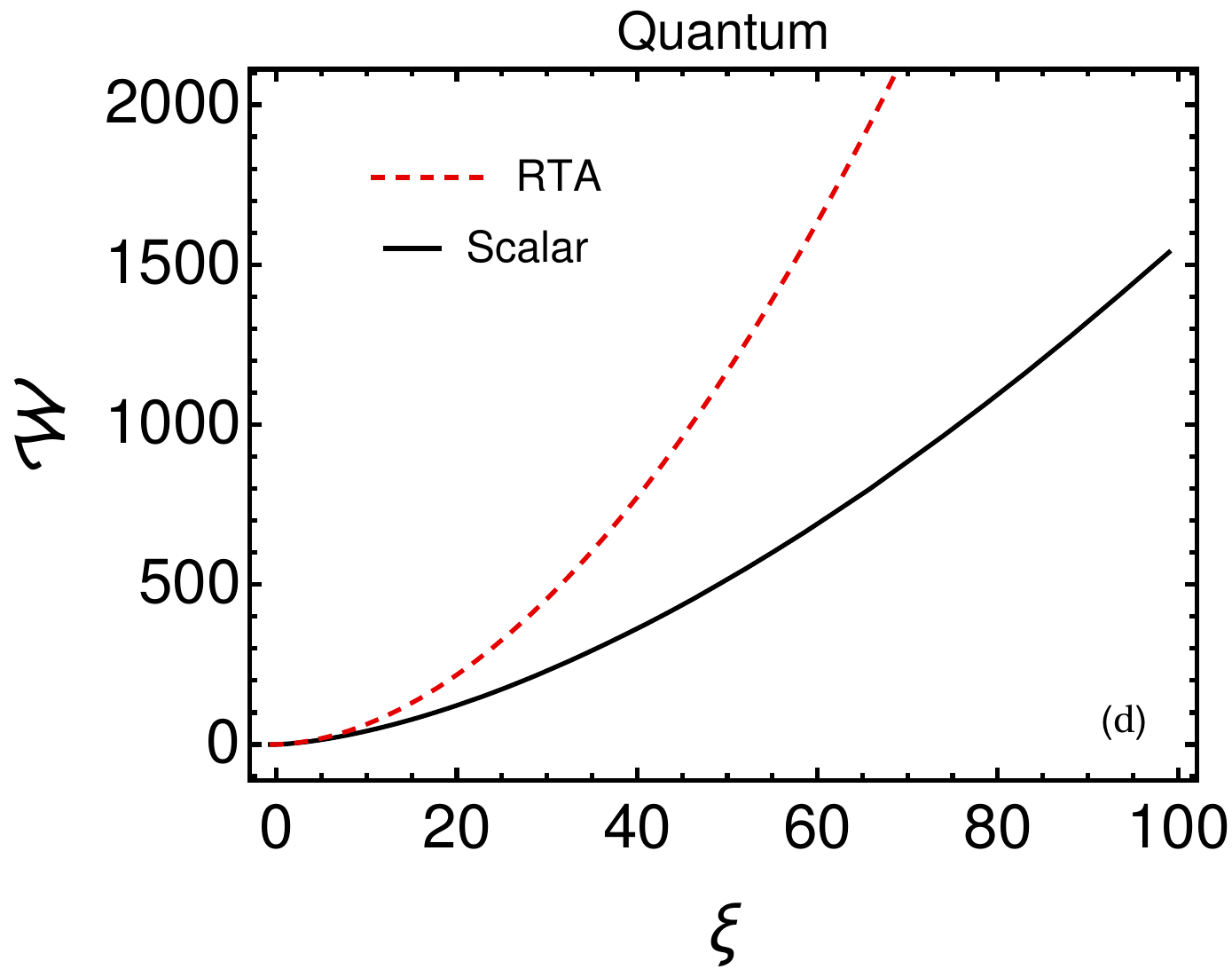}
}
\caption{Comparison of ${\cal W}$ using the LO scalar and RTA kernels.  Panels (a) and (b) in the top row show the classical case ($a=0$) while (c) and (d) from the bottom row show the quantum case ($a=1$).  The left column shows the result for small values of $\xi$ and the right column shows the result for a larger range of values of $\xi$.}
\label{fig:wplots}
\end{figure*}

To proceed, we introduce the special function
\be
{\cal W}(\xi) \equiv \frac{2}{3 \alpha_a {\cal R}^{1/4}(\xi) } \left[ (1+\xi)^{5/2} \tilde{\cal C}^{zz}(\xi) - (1+\xi)^{3/2} \tilde{\cal C}^{xx}(\xi) \right] ,
\label{eq:calw1}
\ee
\checked{m}
to finally write Eq.~(\ref{eq:2mom1}) compactly as
\be
\partial_\tau \xi  - \frac{2(1+\xi)}{\tau} + \frac{{\cal W}(\xi)}{\tau_{\rm eq}}  = 0 \, .
\label{eq:2ndmoment}
\ee
\checked{m}
For future comparisons, note that in RTA one has~\cite{Strickland:2017kux}
\be
{\cal W}(\xi) \rightarrow {\cal W}_{\rm RTA}(\xi) = \xi (1+\xi)^{3/2} {\cal R}^{5/4}(\xi) \, .
\label{eq:calwrta}
\ee
\checked{m}

In Fig.~\ref{fig:wplots} we compare the ${\cal W}$ function obtained from moments of the leading-order scalar and RTA collisional kernels.  The top row shows the classical case ($a=0$) and the bottom row shows the quantum case ($a=1$).\footnote{In RTA, the result is independent of whether one uses classical or quantum statistics.} Panels (a) and (c) show the result for small values of $\xi$ and panels (b) and (d) show the result for a larger range of values of $\xi$.  As can be seen from (a) and (c) of Fig.~\ref{fig:wplots}, for systems that only experience small deviations from equilibrium, the two collisional kernels give very similar results for ${\cal W}$.  However, as (b) and (d) demonstrate, for extremely oblate momentum-space anisotropy, one finds significant differences between the two collisional kernel results for ${\cal W}$.  In all cases shown, we find that ${\cal W}$ obtained with the scalar kernels (classical and quantum) always has a lower magnitude than the RTA kernel.  As a result, one expects to see larger deviations from isotropic equilibrium when using the scalar kernels.

\subsection{Small-$\xi$ limit}

As a check on the result listed above one can take the small-$\xi$ limit using Eq.~(\ref{eq:smallxi3}) and the surrounding discussion to obtain
\be
\lim_{\xi \rightarrow 0} {\cal W} = \xi + {\cal O}(\xi^2) \, ,
\ee
\checked{m}
where we have used the fact that $\lim_{\xi \rightarrow 0} \tilde{\cal C}^{xx} = -\frac{1}{2} \lim_{\xi \rightarrow 0} \tilde{\cal C}^{zz}$.  This agrees with the small-$\xi$ limit of ${\cal W}_{\rm RTA}$.

\section{Numerical solution of the dynamical equations}
\label{sec:numericalsolution}

In this section we present comparisons of the numerical solution of the conformal 0+1d equations of motion obtained in the previous section.  For this purpose, we solve two ordinary differential equations corresponding to Eqs.~(\ref{eq:1stmoment}) and (\ref{eq:2ndmoment}) with some typical initial values for the energy density and pressure anisotropy (shear correction to the pressure) and compare the results obtained using the RTA, classical LO scalar, and quantum LO scalar collisional kernels.

For the scalar collisional kernel we first tabulated 101 points of ${\cal W}(\xi)$ in $-0.68 \leq \xi \leq 99$ using the Monte-Carlo VEGAS with the same parameters/convergence criteria as listed previously.  The resulting numerical data for ${\cal W}(\xi)$ was then fit using a 15$^\text{th}$-order polynomial fit of the form ${\cal W}(\xi) = \sum_{n=0}^{15} c_n \xi^n$.  The resulting fit coefficients for both the classical and quantum cases are listed in Table~\ref{table:coeffs}.  Note that the fact that the linear coefficients are identically one is related to the relaxation-time matching performed between the various collisional kernels.  In addition to this polynomial fit, we performed large-$\xi$ computations and extracted the leading $\xi$-scaling of the kernel in this limit, finding that $\lim_{\xi \rightarrow \infty} {\cal W}(\xi) = w_a \xi^{13/8},$ with $w_0 = 1.1051$ and $w_1 = 0.87962$ for the classical and quantum cases, respectively.  We used the polynomial fit for all $\xi \leq 99$ and the large-$\xi$ result for $\xi > 99$.  The resulting analytic approximations for ${\cal W}(\xi)$ were then used as an input to Eq.~(\ref{eq:2ndmoment}).

\begin{table}[t!]
$
\begin{array}{|c|c|c|}
\hline
\text{\hspace{1cm}} & \text{\bf classical} & \text{\bf quantum} \\
\hline
 c_0 & 0 & 0 \\
 c_1 & 1 & 1 \\
 c_2 & 0.62172 & 0.62101 \\
 c_3 & -0.054309 & -0.082757 \\
 c_4 & 0.0057841 & 0.011445 \\
 c_5 & -0.00044736 & -0.0010753 \\
 c_6 & 0.000024336 & 0.000067954 \\
 c_7 & -9.4415\times 10^{-7} & -2.9582\times 10^{-6} \\
\hline
\end{array}
\hspace{8mm}
\begin{array}{|c|c|c|}
\hline
\text{\hspace{1cm}} & \text{\bf classical} & \text{\bf quantum} \\
\hline

 c_8 & 2.6513\times 10^{-8} & 9.0688\times 10^{-8} \\
 c_9 & -5.4335\times 10^{-10} & -1.9846\times 10^{-9} \\
 c_{10} & 8.122\times 10^{-12} & 3.1112\times 10^{-11} \\
 c_{11} & -8.7563\times 10^{-14} & -3.4641\times 10^{-13} \\
 c_{12} & 6.6303\times 10^{-16} & 2.6729\times 10^{-15} \\
 c_{13} & -3.3466\times 10^{-18} & -1.3581\times 10^{-17} \\
 c_{14} & 1.0114\times 10^{-20} & 4.0854\times 10^{-20} \\
 c_{15} & -1.3852\times 10^{-23} & -5.5098\times 10^{-23} \\
 \hline
\end{array}
$
\caption{Polynomial fit coefficients for the classical and quantum LO scalar ${\cal W}(\xi)$ function defined in Eq.~(\ref{eq:calw1}).  The fit was made assuming ${\cal W}(\xi) = \sum_n c_n \xi^n$ using 101 moment evaluations in the range $-0.68 \leq \xi \leq 99$.}
\label{table:coeffs}
\end{table}

\begin{figure*}[t!]
\centerline{
\includegraphics[width=.49\linewidth]{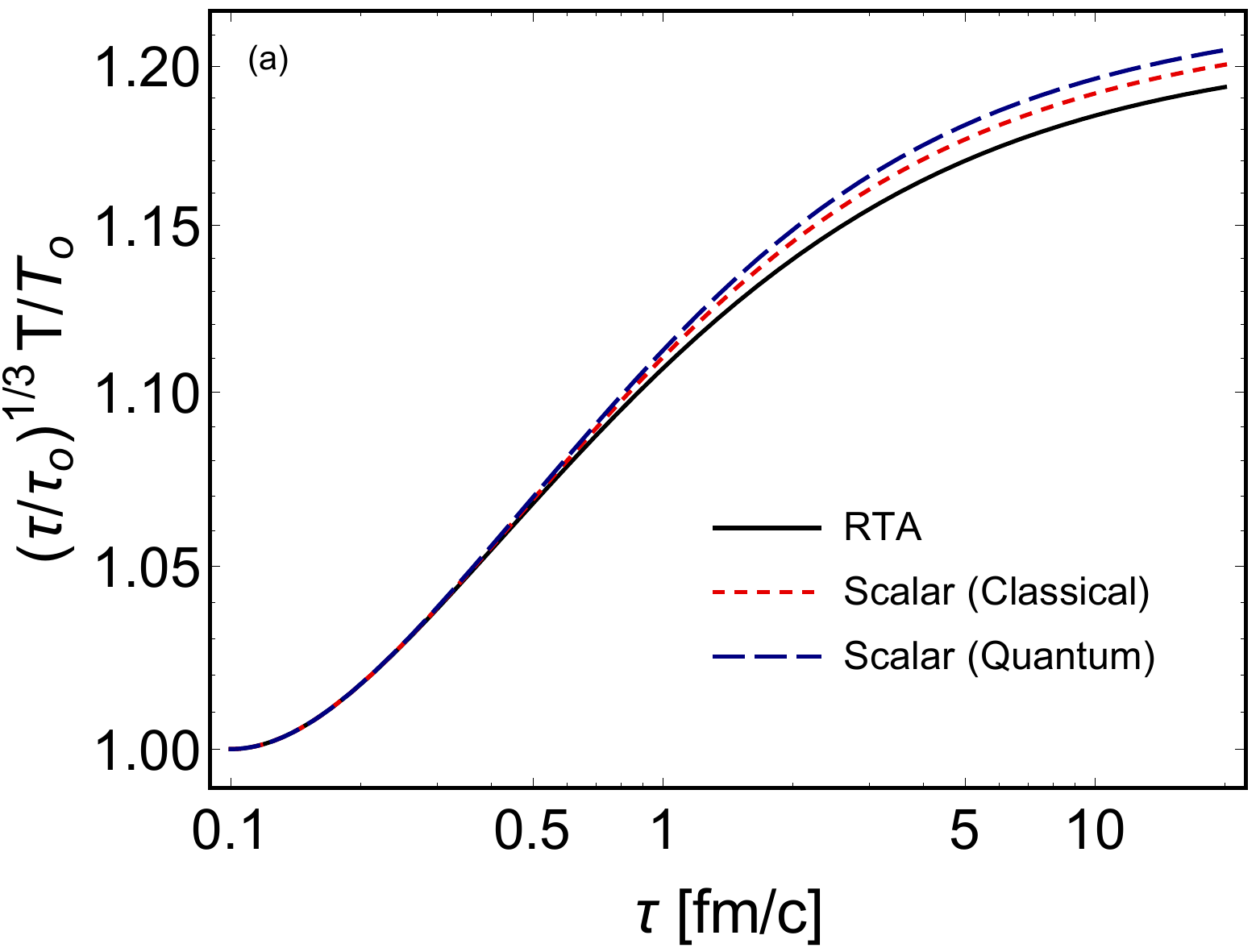}
\hspace{3mm}
\includegraphics[width=.475\linewidth]{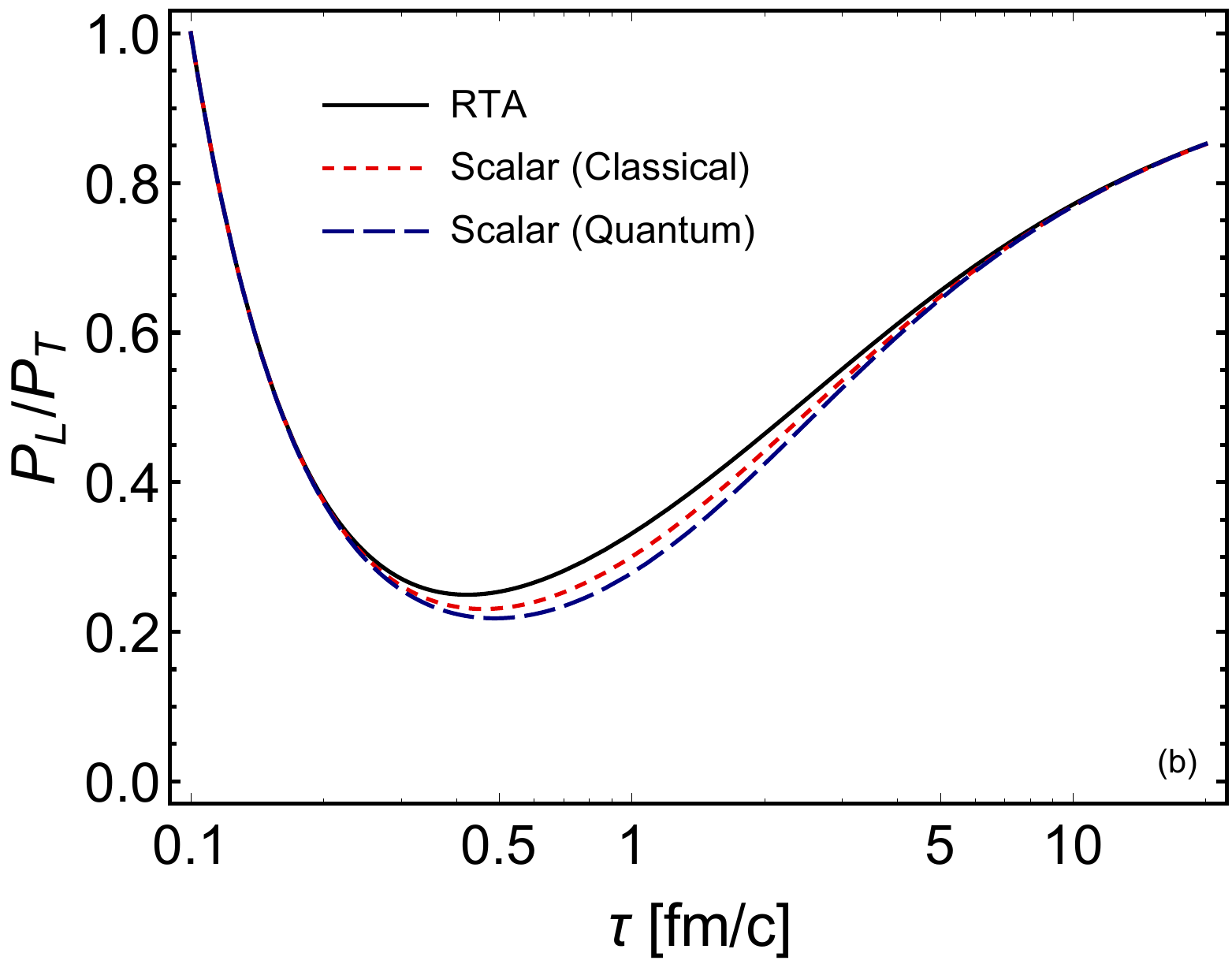}
}
\caption{(Color online) Comparison of the evolution of the scaled temperature (a) and pressure anisotropy (b) for an isotropic initial condition.  The RTA results are indicated by a solid black line, the LO classical scalar results by a short-dashed red line, and the LO quantum scalar results by a long-dashed blue line.}
\label{fig:evolution1}
\end{figure*}

In Fig.~\ref{fig:evolution1} we present results obtained for isotropic initial conditions, ${\cal P}_L(\tau_0)/{\cal P}_T(\tau_0)=1$, with an initial effective temperature of $T_0$ = 500 MeV at $\tau_0$ = 0.1 fm/c using a constant $\bar\eta = 0.2$.  Panel (a) of Fig.~\ref{fig:evolution1} shows the proper-time dependence of the effective temperature $T$ divided by $T_{\rm ideal} = T_0 (\tau_0/\tau)^{1/3}$. Panel (b) of Fig.~\ref{fig:evolution1} shows the pressure anisotropy as a function of proper time.  In both panels, the RTA solutions are indicated by a solid black line, the LO classical scalar result by a short-dashed red line, and the LO quantum scalar result by a long-dashed blue line.  As this figure demonstrates, the effect of the collisional kernel on the temperature evolution is quite small, with the largest deviations occurring a large proper-time.  At $\tau = 20$ fm/c we find that all three results for $T$ are within 1\% of one another.  There is a larger effect on the evolution of the pressure anisotropy, with maximal deviations on the order of 20\% (11\%) between the quantum (classical) LO scalar kernel and RTA at $\tau_0 \simeq 0.7$ fm/c.

\begin{figure*}[t!]
\centerline{
\includegraphics[width=.49\linewidth]{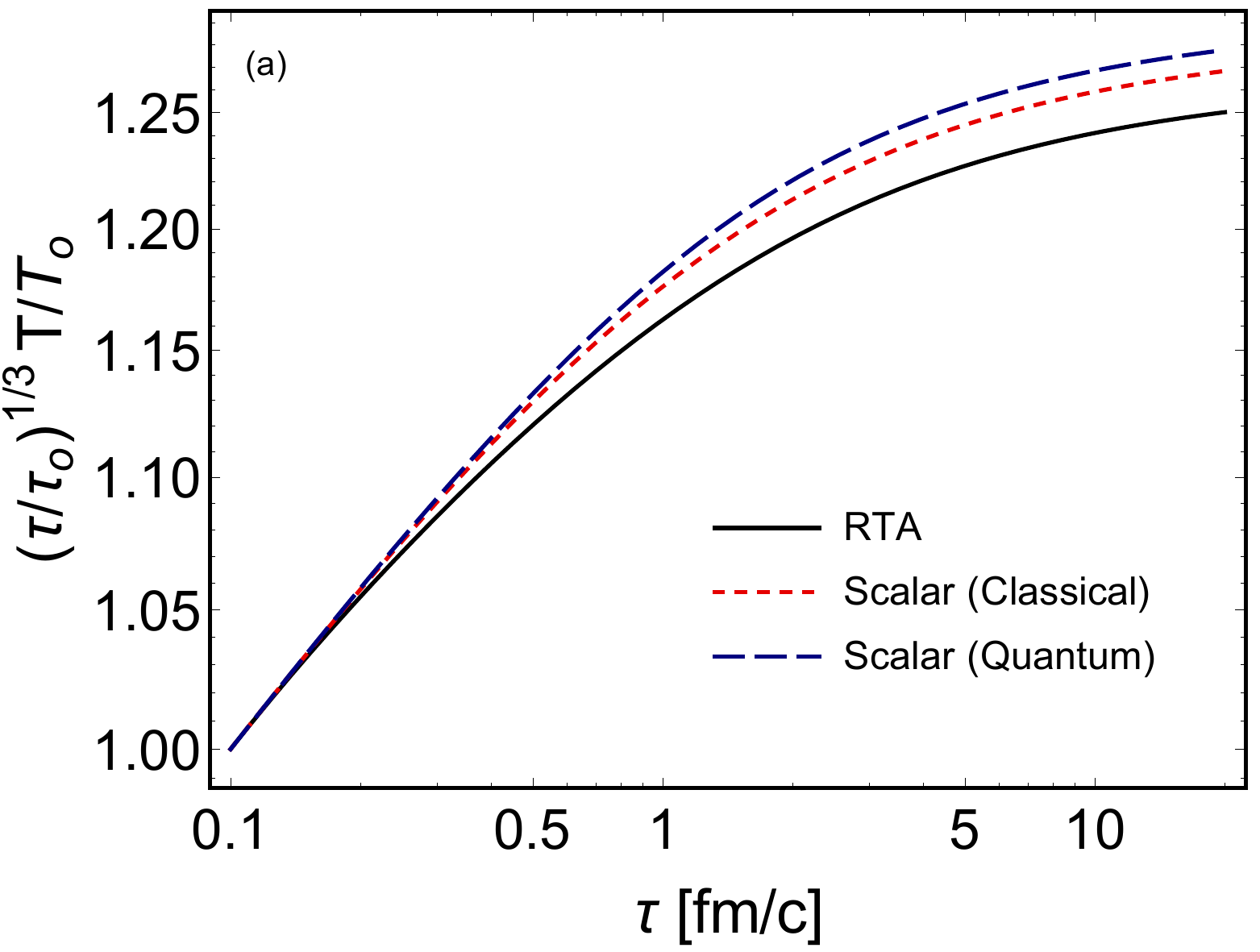}
\hspace{3mm}
\includegraphics[width=.475\linewidth]{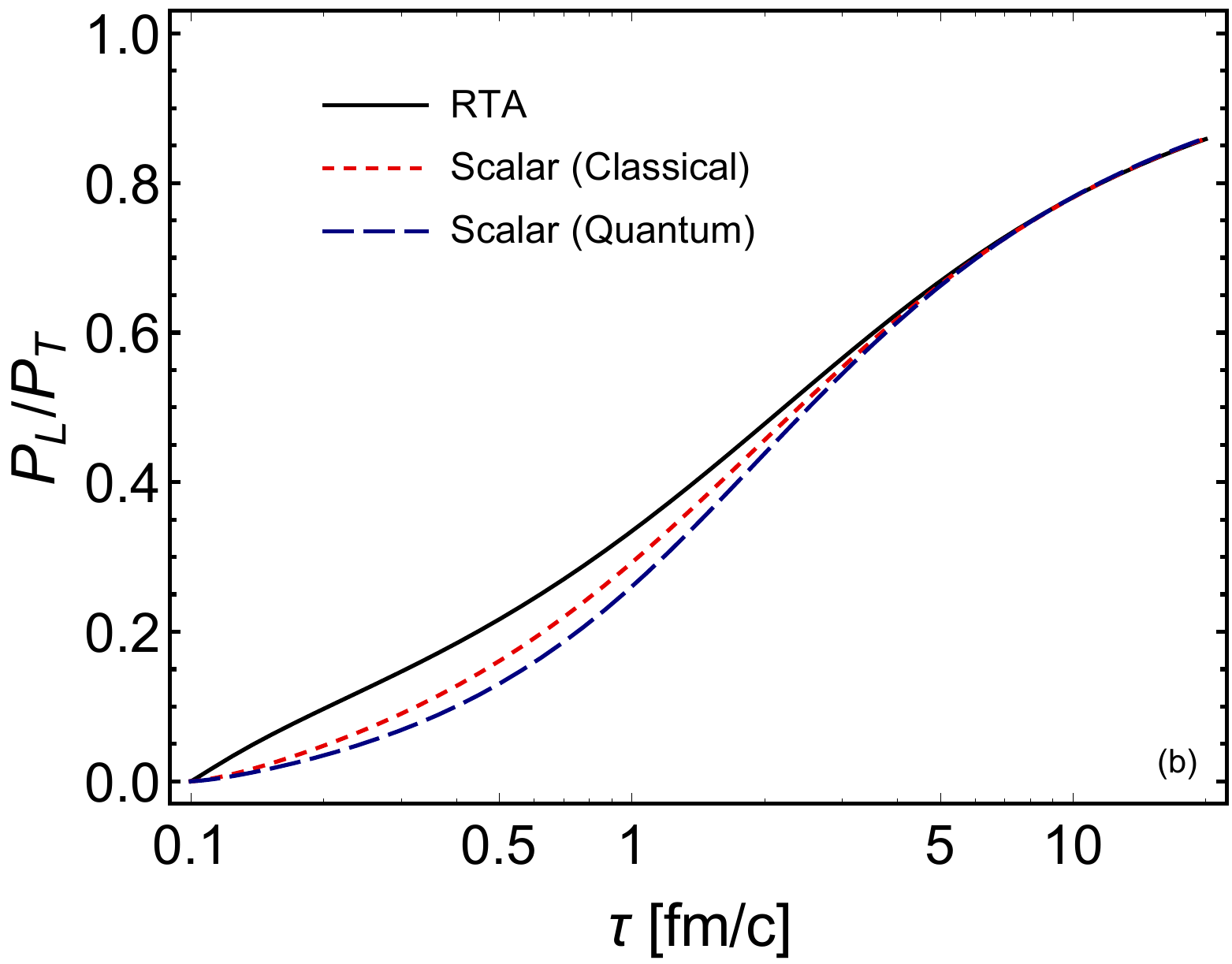}
}
\caption{(Color online) Comparison of the evolution of the scaled temperature (a) and pressure anisotropy (b) for a highly oblate initial condition.  Line styles are the same is in Fig.~\ref{fig:evolution1}.}
\label{fig:evolution2}
\end{figure*}

In Fig.~\ref{fig:evolution2} we present results obtained for oblate initial conditions, ${\cal P}_L(\tau_0)/{\cal P}_T(\tau_0)=10^{-4}$, with an initial effective temperature of $T_0$ = 500 MeV at $\tau_0$ = 0.1 fm/c using a constant $\bar\eta = 0.2$.  The panels and line types are the same as in Fig.~\ref{fig:evolution1}.  As this figure demonstrates, the effect of the collisional kernel on the temperature evolution is once again quite small with the largest deviations occurring at large proper-time.  At $\tau = 20$ fm/c we find that all three results for $T$ are within 2\% of one another.  In this case, however, there is a very large effect on the evolution of the pressure anisotropy, with deviations on the order of 300\% (200\%) between the quantum (classical) LO scalar kernel and RTA at $\tau = 0.2$ fm/c.  We note in closing that the differences in the LO scalar and RTA evolution become larger as one increases $\bar\eta$.

\section{The anisotropic attractor}
\label{sec:attractor}

As we can see from the evolution of ${\cal P}_L/{\cal P}_T$ shown in Fig.~\ref{fig:evolution1}, even if the system is initialized in an isotropic state, it develops a high degree of early-time momentum-space anisotropy due to the rapid longitudinal expansion of the system.  One finds, however, that despite these large momentum-space anisotropies the system is well-described by relativistic dissipative hydrodynamics.  The timescale for the onset of dissipative hydrodynamical behavior in the QGP has been dubbed the``hydrodynamization'' time scale and researchers have found that this time scale is generically much shorter than the isotropization time scale~\cite{Chesler:2008hg,Beuf:2009cx,Chesler:2009cy,Heller:2011ju,Heller:2012je,Heller:2012km,vanderSchee:2012qj,Casalderrey-Solana:2013aba,vanderSchee:2013pia,Heller:2013oxa,Keegan:2015avk,Chesler:2015bba,Kurkela:2015qoa,Chesler:2016ceu,Attems:2016ugt,Attems:2016tby,Attems:2017zam,Florkowski:2017olj,Alqahtani:2017mhy}.  

Recently, it has been demonstrated that the process of hydrodynamization is driven by a non-equilibrium dynamical attractor; the details of which depend on the specific theory/model under consideration~\cite{Heller:2015dha,Keegan:2015avk,Florkowski:2017olj,Romatschke:2017vte,Strickland:2017kux,Romatschke:2017acs,Behtash:2017wqg,Denicol:2017lxn,Alqahtani:2017mhy}.  In a recent paper \cite{Strickland:2017kux} it was shown how to determine the dynamical attractor associated with aHydro and two different second-order vHydro frameworks: DNMR and Mueller-Israel-Stewart (MIS); however, in both cases an RTA collisional kernel was assumed.  In this section, we would like to present the first results for the aHydro attractor using a LO scalar collisional kernel.  To determine the attractor, one introduces new variables, which are the scaled proper-time $w \equiv \tau T(\tau)$ and the amplitude $\varphi(w)$ defined as~\cite{Heller:2015dha,Strickland:2017kux} 
\be
\varphi(w) \equiv \tau \frac{\dot w}{w} = 1 + \frac{\tau}{4} \partial_\tau\!\log \varepsilon \, .
\label{wdot1}
\ee
\checked{md}
The amplitude $\varphi$ is related to the single independent component of the shear-stress tensor $\pi \equiv \pi^\eta_\eta$ in 0+1d as follows
\be
\bar\pi \equiv \frac{\pi}{\varepsilon} = 4\left(\varphi -\frac{2}{3}\right) ,
\ee 
\checked{md}
and the pressures are given by $P_L = P_{\rm eq} + \pi$ and $P_T =  P_{\rm eq} - \pi/2$.

\begin{figure*}[t!]
\centerline{
\includegraphics[width=.475\linewidth]{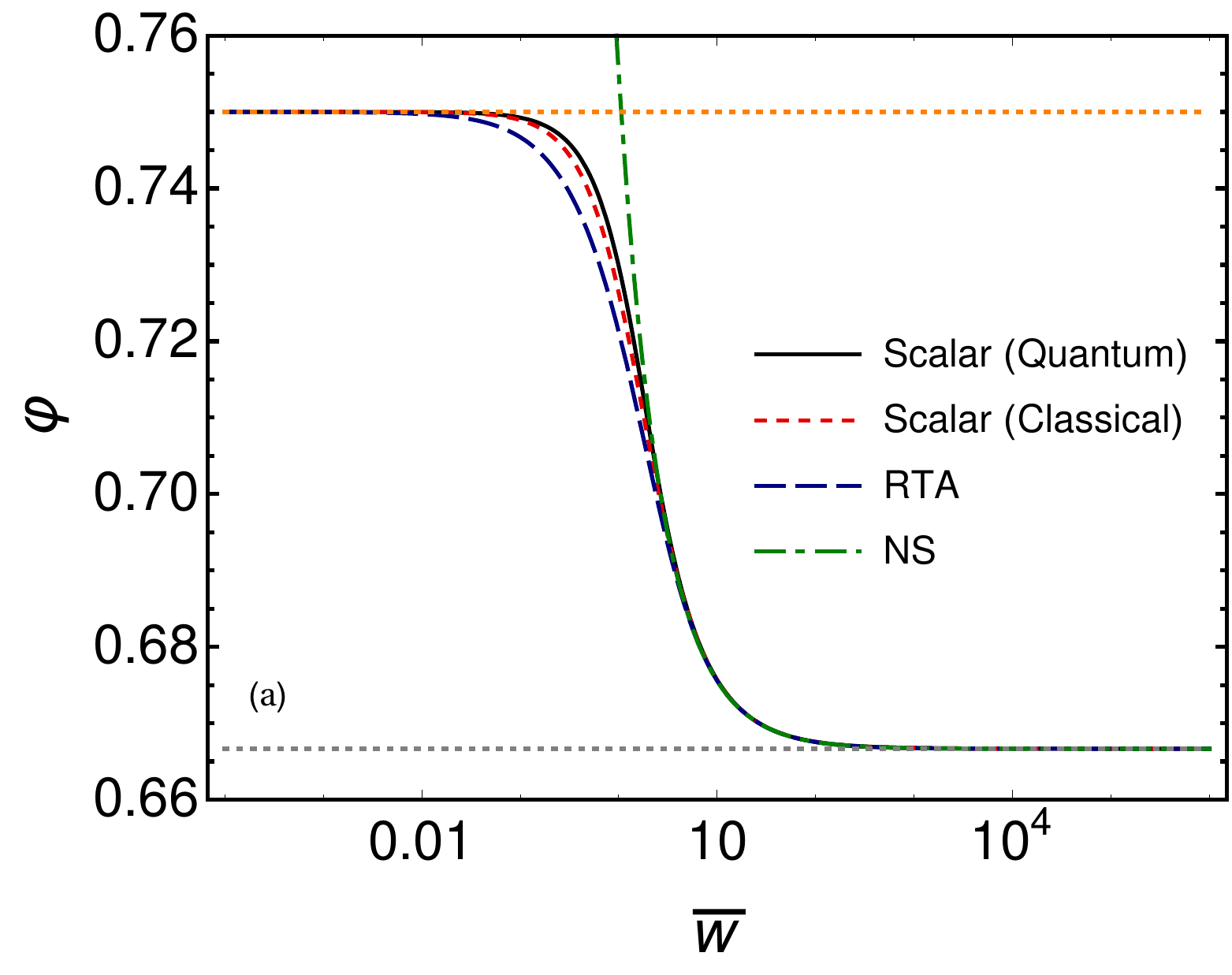}
\hspace{3mm}
\includegraphics[width=.475\linewidth]{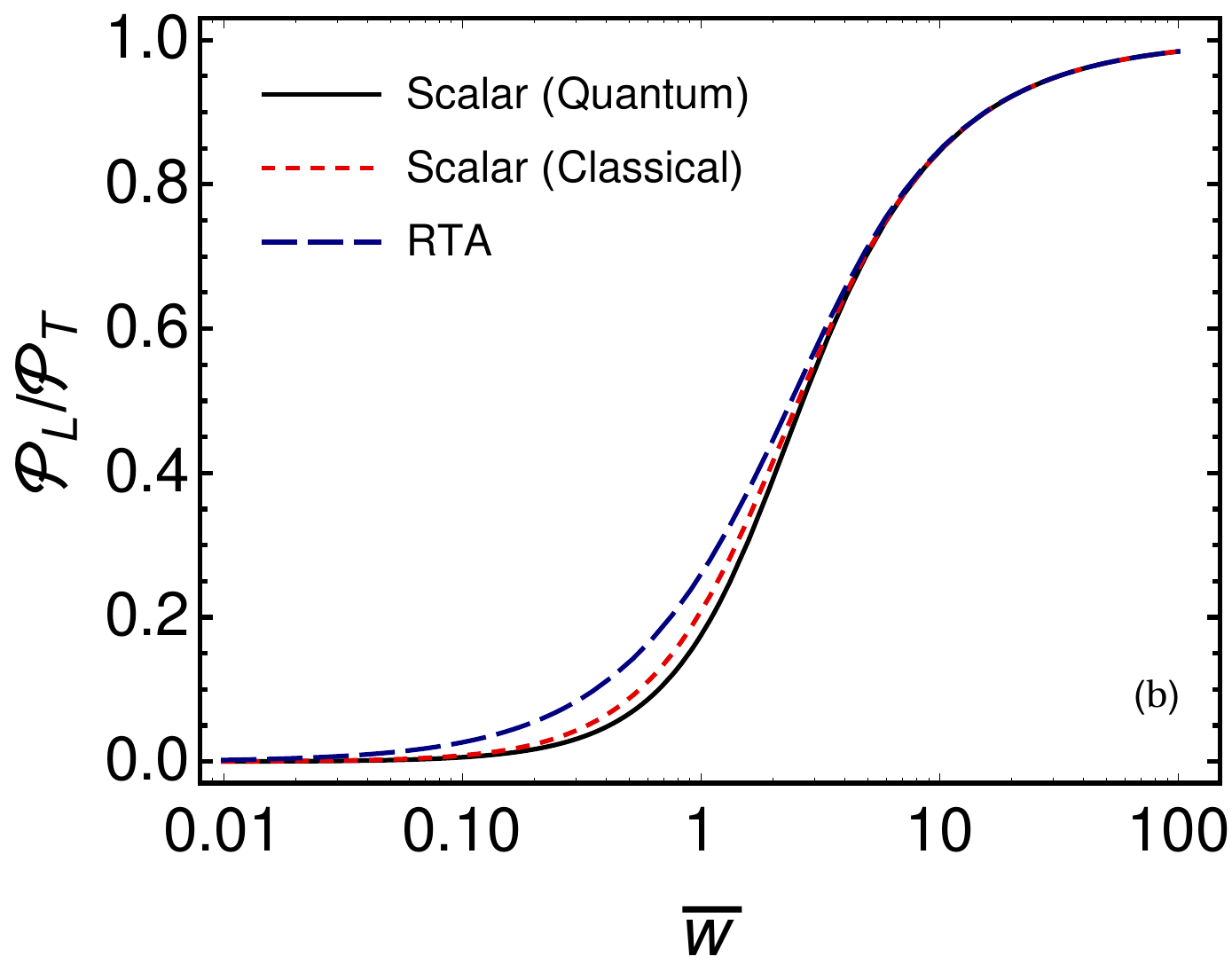}
}
\vspace{-4mm}
\caption{(Color online) Panel (a) shows the attractor amplitude $\varphi$ as a function of $\overline{w}$ andpanel (b) shows the pressure anisotropy ${\cal P}_L/{\cal P}_T$ as a function of $\overline{w}$.  In both panels, the scalar quantum result is indicated by a solid black line, the scalar classical result is indicated by a red short-dashed line, and the RTA result is indicated by a blue long-dashed line.  In panel (a), we additionally show the Navier-Stokes (NS) result as a green dot-dashed line and the asymptotic bounds $\varphi = 2/3$ and $\varphi=3/4$ as grey and orange dotted lines, respectively.}
\label{fig:attractor}
\end{figure*}

Using the method detailed in Ref.~\cite{Strickland:2017kux} one finds the following differential equation for the aHydro attractor with the leading-order scalar scattering kernel
\be
\overline{w} { \varphi} \frac{\partial \varphi}{\partial \overline{w}}  = \left[ \frac{1}{2} (1+\xi) - \frac{\overline{w}}{4} {\cal W}(\xi) \right] \bar\pi' \, .
\label{eq:ahydroattractoreq1}
\ee
\checked{md}
where $\overline{w} \equiv w/c_\pi$ with $c_\pi \equiv \tau_{\rm eq} T = 5 \bar\eta$, ${\cal W}$ defined in Eq.~(\ref{eq:calw1}), and $\bar\pi'(\varphi)$ being the first-derivative of $\bar\pi$ with respect to $\xi$.\footnote{In Ref.~\cite{Strickland:2017kux} the ${\cal W}$ function was called ${\cal H}$.  We have renamed it to avoid any possible confusion with existing ${\cal H}$ functions in the aHydro framework.}  In all cases, $\xi$ is understood to be evaluated using the nonlinear inverse function which relates $\varphi$ and $\xi$~\cite{Strickland:2017kux}.  We will compare solutions to Eq.~(\ref{eq:ahydroattractoreq1}) using the RTA collisions kernel for which ${\cal W}$ is given by Eq.~(\ref{eq:calwrta}).  We will also compare with the Navier-Stokes result~\cite{Heller:2015dha} 
\be
\varphi_{\rm NS} = \frac{2}{3} + \frac{4}{9} \frac{c_{\eta/\pi}}{\overline{w}} \, .
\ee
\checked{md}
In RTA, one has $c_{\eta/\pi} = 5$.

\begin{figure*}[t!]
\centerline{
\includegraphics[width=.475\linewidth]{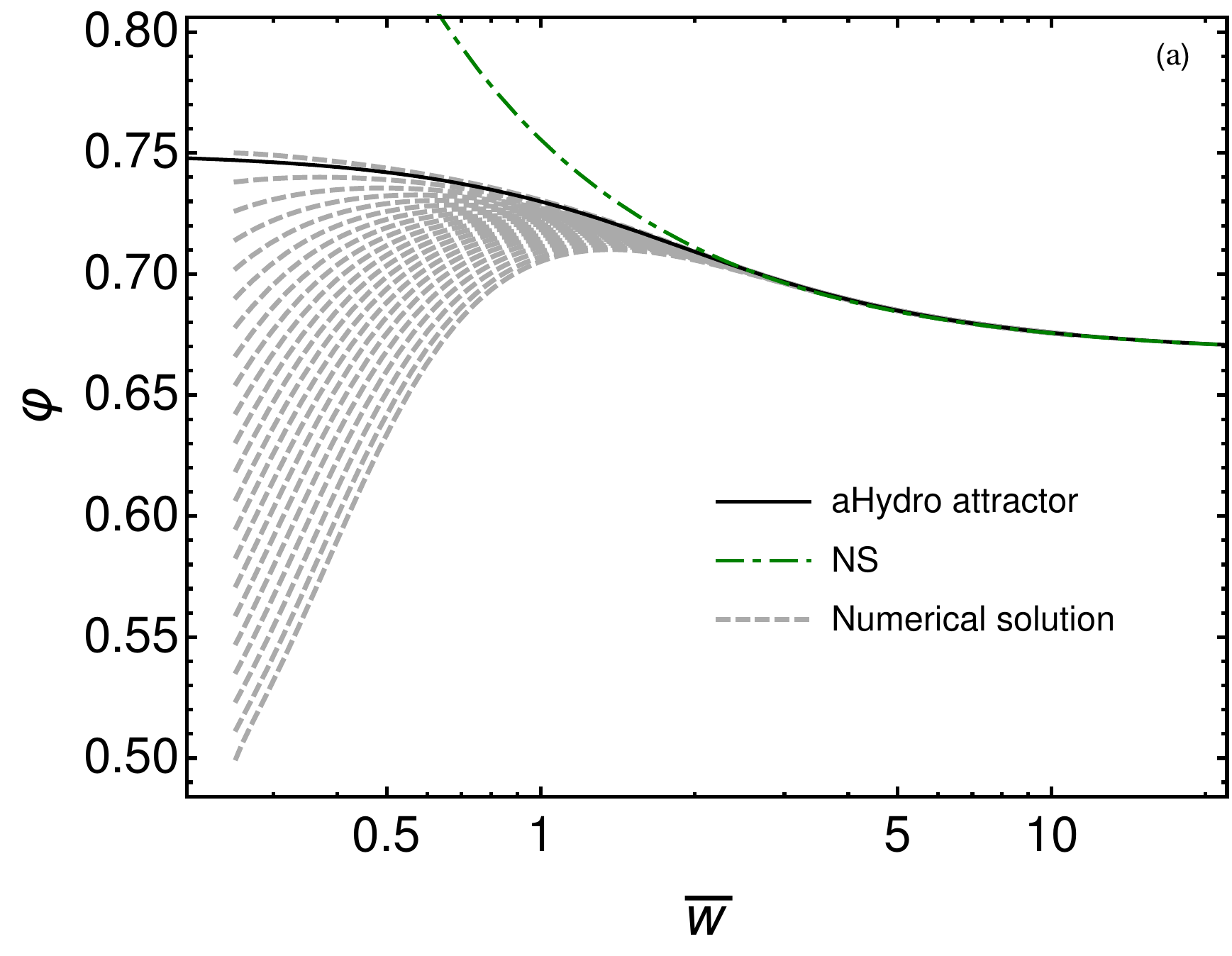}
\hspace{3mm}
\includegraphics[width=.475\linewidth]{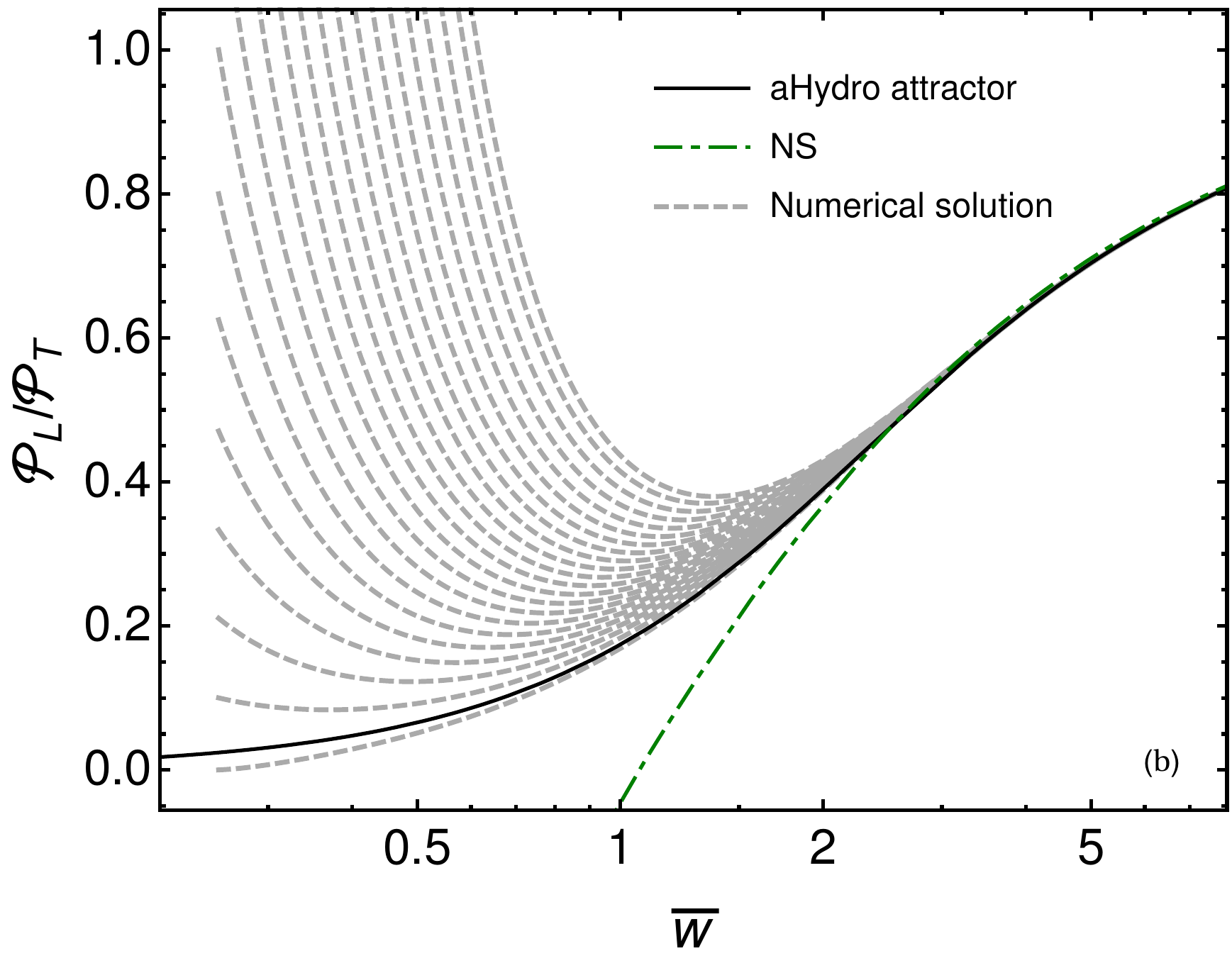}
}
\vspace{-4mm}
\caption{(Color online)  Comparison of the classical LO scalar attractor with a set of numerical solutions to the equations of motion for a variety of initial conditions.  Panel (a) shows $\varphi$ and panel (b) shows the resulting pressure anisotropy.}
\label{fig:sols1}
\end{figure*}
\begin{figure*}[t!]
\centerline{
\includegraphics[width=.475\linewidth]{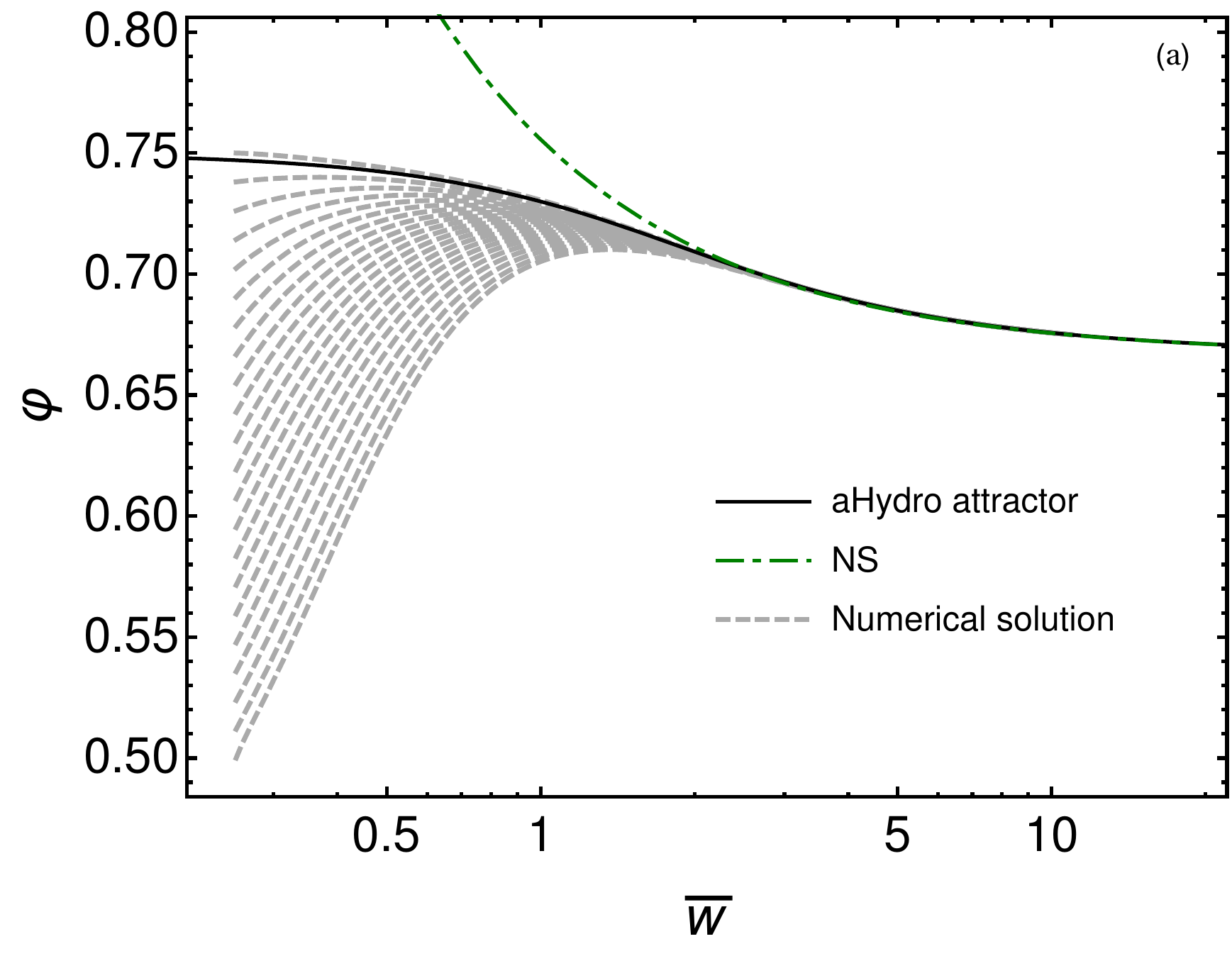}
\hspace{3mm}
\includegraphics[width=.475\linewidth]{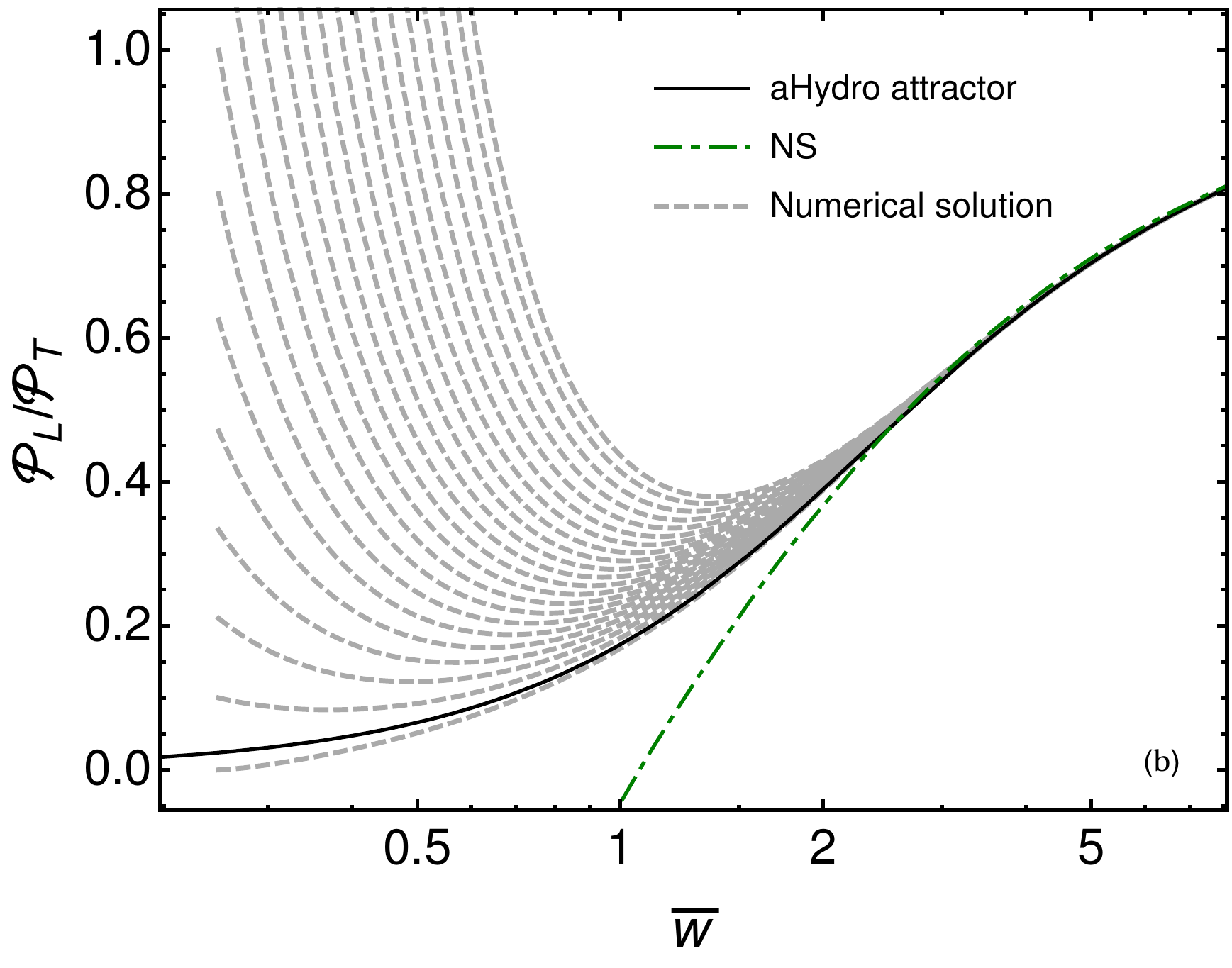}
}
\vspace{-4mm}
\caption{(Color online)  Comparison of the quantum LO scalar attractor with a set of numerical solutions to the equations of motion for a variety of initial conditions.  Panel (a) shows $\varphi$ and panel (b) shows the resulting pressure anisotropy.}
\label{fig:sols2}
\end{figure*}

In Fig.~\ref{fig:attractor} panel (a) shows the attractor amplitude $\varphi$ as a function of $\overline{w}$ and panel (b) shows the pressure anisotropy ${\cal P}_L/{\cal P}_T$ as a function of $\overline{w}$.  As we can see from this figure, the RTA and scalar attractors are quantitatively different but share many qualitative features, e.g. bounding values, width of the transition region, etc.  From the pressure anisotropy plot (right panel), we see that both scalar collisional kernels result in an attractor which possesses a higher degree of momentum anisotropy, consistent with the specific solutions presented in the previous section.  In Figs.~\ref{fig:sols1} and \ref{fig:sols2} we show a comparison of the classical LO scalar attractor with a set of numerical solutions to the equations of motion for a variety of initial conditions.  The grey-dashed lines were generated by varying the initial $\varphi$ in the range $1/2 \leq \varphi \leq 3/4$ with the bounds corresponding to infinitely prolate and oblate initial conditions, respectively.  As these figures demonstrate, both the classical and quantum evolutions rapidly approach the attractor solution for their respective cases.  We see no qualitative difference between the approach of the solutions in the case of the LO scalar collisional kernel and that found for the case of RTA in Ref.~\cite{Strickland:2017kux}.  In Ref.~\cite{Strickland:2017kux} it was demonstrated that the aHydro RTA attractor was virtually indistinguishable from the exact RTA attractor determined by iterative solution of the RTA Boltzmann equation \cite{Florkowski:2013lza,Florkowski:2013lya}.  For this reason, we expect that the aHydro attractor determined using the LO scalar collisional kernel would be a very good approximation to the corresponding exact kinetic attractor with this kernel.  

Although our Figs.~\ref{fig:sols1} and \ref{fig:sols2} and Fig.~6 from Ref.~\cite{Strickland:2017kux} show that the approach to each kernel's respective attractor is qualitatively the same in all cases considered, quantitative differences remain in the rate of approach to the attractor.  In order to quantify the different rates of approach to the attractor solution, we have numerically extracted the leading asymptotic behavior of a generic solution by measuring the ``damping coefficient'' $\gamma$ defined via
\be
\frac{{\cal P}_L}{{\cal P}_T} - \left( \frac{{\cal P}_L}{{\cal P}_T} \right)_{\rm attractor} \simeq A e^{-\gamma \overline{w}} \, ,
\ee
at large $\overline{w}$.  In practice, we made fits in the region $2 \leq \overline{w} \leq 4$ where this behavior was clearly observed and averaged over the set of initial conditions shown in our Figs.~\ref{fig:sols1} and \ref{fig:sols2} in order to extract the logarithmic slope and intercept using a least-squares fit.  We find that $\gamma_0 = 1.73 \pm 0.01$,  $\gamma_1 = 1.63 \pm 0.01$, and $\gamma_{\rm RTA} = 1.88 \pm 0.01$, for the classical scalar, quantum scalar, and RTA cases, respectively.  This indicates that the approach to the non-equilibrium attractor is fastest for RTA and slowest for the quantum scalar collisional kernel.  This conclusion is further evidenced by measuring the value of $\overline{w}$ necessary for all solutions shown in Figs.~\ref{fig:sols1} and \ref{fig:sols2} and Fig.~6 from Ref.~\cite{Strickland:2017kux} to come within 1\% of their respective attractor solutions.  We find $\overline{w}_{1\%} = \left\{ 2.88,3.19,2.68 \right\} \pm 0.01$ for the classical scalar, quantum scalar, and RTA kernels, respectively, indicating again that the RTA kernel dynamics approaches its attractor most quickly and the quantum scalar kernel most slowly.

\section{Conclusions and outlook}
\label{sec:conclusions}

In this paper we presented first results of using a more realistic collisional kernel in the context of anisotropic hydrodynamics.  This is a step forward from prior works, which have all used the RTA collisional kernel or some variant thereof.  We demonstrated that in order to use a general $2 \leftrightarrow 2$ scattering kernel, one can reduce the problem to computing a finite set of eight-dimensional integrals as a function of one or more anisotropy parameters.  In the specific case of conformal 0+1d Bjorken expansion, we demonstrated that one only needs to tabulate two moments ${\cal C}^{xx}$ and ${\cal C}^{zz}$ as a function of a single anisotropy parameter $\xi$.  Herein, we did this numerically for LO scalar $\lambda \phi^4$ theory by evaluating the required moments using Monte-Carlo VEGAS integration.  The numerical results determined in this manner were then combined into a single function ${\cal W}(\xi)$ (\ref{eq:calw1}) which contains all information about the collisional kernel necessary to obtain and solve the equations of motion.

To further simplify the result, we tabulated ${\cal W}(\xi)$ on a grid in $\xi$, made a polynomial fit with the resulting classical and quantum coefficients listed in Table \ref{table:coeffs}, and additionally performed large-$\xi$ expansions in both cases. The resulting approximations will allow anyone to study the effect of the scalar collisional kernel without having to perform the eight-dimensional Monte-Carlo integrations on their own.  Comparing the evolution obtained using RTA and the LO scalar $\lambda \phi^4$ collisional kernel we find that, when the relaxation times are matched in the near-equilibrium limit, one finds, that for a given fixed value of $\bar\eta$, the temperature evolution is the same to within a few percent, however, the pressure anisotropy developed is higher with the scalar kernel than with RTA.  The differences in evolution were found to be larger when the initial momentum-space anisotropy was large or the shear viscosity to entropy density ratio was large.  The conclusion that the pressure anisotropy is larger when using both the classical and quantum scalar kernels was further evidenced by studying the dynamical attractor associated with the LO scalar kernel, where it was found that the LO scalar attractors possessed a higher degree of momentum-space anisotropy than the RTA attractor.  Additionally, we demonstrated that the rate of approach to each kernel's respective dynamical attractor is quantitatively different, with the RTA kernel resulting in the fastest approach and quantum scalar kernel resulting in the slowest approach among the three cases considered herein.

Looking forward, the work presented here lays the groundwork for the use of more realistic QCD-based collisional kernels in the context of aHydro.  In particular, one can use the effective kinetic theory collisional kernel from Ref.~\cite{Kurkela:2015qoa} which self-consistently includes both elastic and inelastic gluon scattering.  Work along these lines is in progress \cite{forth}.

\acknowledgments{
D.\ Almaalol was supported by a fellowship from the University of Zawia.  M.\ Strickland was supported by the U.S. Department of Energy, Office of Science, Office of Nuclear Physics under Award No.~DE-SC0013470.
}

\bibliography{kernel}

\end{document}